\documentclass{aa}

\usepackage{graphicx}
\usepackage{txfonts}
\usepackage{caption}
\usepackage{subcaption}
\usepackage{amsmath,amsfonts,bm}
\usepackage{comment}
\usepackage{cleveref}
\begin{document} 

   \title{Total solar eclipse 2024 modelling with COCONUT}


   \author{T. Baratashvili \inst{1}, H. P. Wang \inst{1}, D. Sorokina \inst{1}, A. Lani\inst{1}, S. Poedts \inst{1,2}
          }

   \institute{Department of Mathematics/Centre for mathematical Plasma Astrophysics, 
             KU Leuven, Celestijnenlaan 200B, 3001 Leuven, Belgium. 
             \email{tinatin.baratashvili@kuleuven.be}
             \and
             Institute of Physics, University of Maria Curie-Sk{\l}odowska, 
             ul.\ Radziszewskiego 10, 20-031 Lublin, Poland}
             
\date{Accepted: October 30, 2025}
 
    \abstract
   {Coronal modelling is crucial for a better understanding of solar and helio-physics. Accurate plasma conditions at the outer boundary of a solar corona lead to more precise space weather predictions. Due to the strong brightness of the Sun and the lack of white light observations of the solar atmosphere and low corona (1-1.5R$_\odot$), total solar eclipses have become a standard approach for validating the coronal models. In this study, we validate the COCONUT coronal model by predicting the coronal configuration during the total solar eclipse on April 8, 2024. }  
   {We aim to predict the accurate configuration of the solar corona during the total solar eclipse on April 8, 2024. Additionally, we compare the predictive capabilities in the steady and dynamic driving of the boundary conditions in COCONUT. We utilise the full 3D MHD model to reconstruct the solar corona from the solar surface to $30\;R_\odot$.}
   {We started predicting the upcoming total solar eclipse on March 21. The predictions were conducted in three different regimes: quasi-steady driving of the inner boundary conditions (BCs) with a daily cadence and dynamic driving of the inner BCs with both daily and hourly cadences. The results from all the simulations are compared to the total solar eclipse images. Additionally, the synthetic white-light (WL) images are generated from the Solar Terrestrial Relations Observatory Ahead (STEREO-A) field of view and compared to COR2 observed images. Normalised polarised brightness is compared in the COR2 and synthetic WL images.}
   {The predicted solar corona does not vary significantly in the first half of the prediction window, since the observations of the magnetic field are not yet accurately updated on the solar surface corresponding to April 8. The magnetic field changes more significantly as the eclipse date approaches. The dynamic simulations yielded better results than the quasi-steady predictions. Driving the simulations at a higher cadence did not significantly improve the results, due to the highly processed magnetic field maps. The west limb was reconstructed better in the simulations than the east limb. }
   {We have predicted the total solar eclipse coronal configuration 18 days before the total solar eclipse. We can conclude that the dynamic simulations produced more accurate predictions. The availability of comprehensive observations is crucial, as the emergence of the active region on the east limb made it difficult to accurately predict the east limb coronal configuration due to incorrect input of magnetic field data. More detailed input BCs are necessary for reconstructing smaller-scale structures in the corona. }

   \keywords{}
\titlerunning{Eclipse 2024 modelling with COCONUT}
\authorrunning{Baratashvili et al.}
   \maketitle
%

\section{Introduction} \label{section:introduction}
Complex physical processes take place in the solar corona. These processes are not fully understood since observing them is difficult or even impossible. Hence, there is a significant lack of available data to test different theories and validate models. The solar corona is faint compared to the luminosity coming from the solar surface. Therefore, it is impossible to distinguish the details that characterise different phenomena in the solar corona. The standard approach has become to test various theories about the processes in the solar atmosphere with coronal modelling. Validating the modelled corona is crucial in advancing our knowledge and understanding of solar physics. Since we lack white light coronagraph observations in the low solar corona 1-1.5R$_\odot$, total solar eclipses have become a standard way of assessing the performance of coronal models. Total solar eclipses constrain the validation period of coronal modelling, as they occur infrequently. However, currently it is the only way of distinguishing the features in the low corona close to the solar surface. Total solar eclipses represent a single snapshot of the solar coronal configuration in time, while for time-dependent simulations, continuous white-light observations are necessary. In the near future, the PROBA-3 mission will provide us with continuous white-light (WL) images of the corona, starting from near the solar surface \citep{Zhukov2025}. Still, to this day, pictures taken during total solar eclipses remain one of the most popular ways of validating coronal models. 

A total solar eclipse occurred on April 8, 2024, during the maximum of solar cycle 25. Therefore, the eclipse images revealed a very complex structure of the solar corona, with multiple streamers and a prominence visible on the south-west limb of the Sun. The total solar eclipse of 2024 presents an excellent opportunity to test the performance of coronal models; however, it is also very challenging due to the strong magnetic fields and gradients that occur during solar maximum. 

Numerous global coronal models have been developed over the years. One of the oldest and most mature numerical models is magnetohydrodynamics (MHD) around a sphere (MAS) by Predictive Science Inc. \citep{Mikic1999,MikicLinker1996}. The team has advanced the model, starting from the total solar eclipse in 1996, and has shown increasingly accurate predictions since the late 1990s. {\cite{Downs2025} reports the results of modelling the same eclipse event performed by the MAS model.} Another advanced global MHD model is the AWSoM model \citep{Vanderholst2014}, which uses the Alfv\'en wave heating mechanism for solar wind acceleration. Other models, such as MULTI-VP \citep{Pinto2017}, SIP-IFVM \citep{wang2025sipifvmtimeevolvingcoronalmodel}, Wind-Predict \citep{reville2016, reville2020}, and a 3D MHD model developed by \cite{Usmanov2014, Usmanov20189}, also focus on global coronal structures using various approaches. Numerical computations in the coronal domain are relatively expensive and complex. The outcome heavily depends on the imposed boundary conditions and the implemented physics and numerical techniques. 

Recently, a new, optimised 3D global coronal modelling tool COolfluid COroNa UnsTructured (COCONUT) \citep{Perri2022}, was developed within the framework of the Computational Object-Oriented Libraries for Fluid Dynamics \citep[COOLFluiD]{Lani2005, Kimpe2005, Lani2006, Lani2013, Lani2014} platform. The first version of COCONUT was polytropic and primarily focused on the inner corona structures near the Sun. Later, \cite{Baratashvili2024C} upgraded the polytropic COCONUT model to a full MHD model, including source and sink terms in the equations to accelerate and decelerate the solar wind to obtain a bi-modal solar wind velocity distribution. This upgraded COCONUT model is suitable for space weather purposes, as the output of the corona model was successfully coupled to the heliospheric models Icarus \citep{Baratashvili2025B} and EUHFORIA \citep{Pomoell2018}. This upgraded version of COCONUT is constrained by static magnetograms that are used as inner boundary conditions for the radial magnetic field component. Moreover, \cite{Wang2025} and \cite{wang2025COCONUTMayEvent} extended COCONUT into a more realistic, time-evolving coronal model driven by a series of subsequent magnetograms that are interpolated to generate time-dependent boundary conditions.

This paper uses the 2024 total solar eclipse to validate the novel global 3D MHD coronal model, COCONUT. The predictions began on March 21, 2024, in a quasi-steady state {regime} and were updated daily with the latest magnetogram data from the GONG observatory. After the total solar eclipse, we used the observed photospheric magnetic field closest to the eclipse time to drive the coronal model. Additionally, we performed the simulations in the time-dependent regime, with daily and hourly updated magnetic field information at the inner boundary. 
Section~\ref{sec:coconut_setup} describes the numerical setup in COCONUT, while section~\ref{sec:input_magnetograms} focuses on the input data for various simulations. The predicted solar corona and multiple validation techniques are described in section~\ref{sec:predictions} for all the simulation setups. Finally, we present our conclusions in section~\ref{sec:conclusions}.

\section{Numerical setup in COCONUT} \label{sec:coconut_setup}

COCONUT is a 3D MHD coronal model with a domain starting from near the solar surface to $30\;$R$_\odot$. The data-driven model uses the observed photospheric magnetic field to model its corona. We solve the 3D MHD equations in their conservative form, reading:
\begin{multline}
\frac{\partial}{\partial t}\left(\begin{array}{c}
\rho \\
\rho \vec{V} \\
\vec{B} \\
E \\
\phi
\end{array}\right)+\vec{\nabla} \cdot \left(\begin{array}{c}
\rho \vec{V} \\
\rho \vec{V} \vec{V}+\tens I\left(p+\frac{1}{2}|\vec{B}|^{2}\right)-\vec{B} \vec{B} \\
\vec{V} \vec{B}-\vec{B} \vec{V}+\underline{\tens I \phi} \\
\left(E+p+\frac{1}{2}|\vec{B}|^{2}\right) \vec{V}-\vec{B}(\vec{V} \cdot \vec{B}) \\
V^2_\text{ref}\mathbf{B}
\end{array}\right) \\ =\left(\begin{array}{c}
0 \\
\rho \vec{g}\\
0 \\
\rho \vec{g} \cdot \vec{V} + \mathbf{S} \\
0
\end{array}\right),
\end{multline}
where ${E}$ denotes the total energy $\rho \frac{V^2}{2} + \rho \mathcal E + \frac{B^2}{8\pi}$, $\vec{B}$ is the magnetic field, $\vec{V}$ the velocity field, $\vec{g}$ the gravitational acceleration, $\rho$ the plasma density, and $p$ the thermal pressure. The gravitational acceleration is given by $\vec{g}(r) = -(G M_\odot/r^2)\, \hat{\vec{e}}_r$ and the identity dyadic $ \tens I = \hat{\vec{e}}_x \otimes \hat{\vec{e}}_x + \hat{\vec{e}}_y \otimes \hat{\vec{e}}_y + \hat{\vec{e}}_z \otimes \hat{\vec{e}}_z$. The closure is given by the ideal equation of state, thus giving for the internal energy density $\rho \mathcal E = P/(\gamma - 1)$, with an adiabatic index $\gamma$ of $5/3$. The solar rotation is considered by prescribing the $V_\phi$ component at the inner boundary \citep{Kuzma2023}. $\mathbf{S}$ represents the source and sink terms, including thermal conductivity, radiative losses and an ad hoc coronal heating function in the right-hand side of the energy equation,
$\mathbf{S} = - \nabla \cdot \mathbf{q} + Q_{rad} + Q_{H}$. The source and sink terms are described in detail by \cite{Baratashvili2024C}. For the approximated heating function, we choose the following formulation:
\begin{equation} \label{magnetic_damping}
    Q_{H} = H_0 \cdot |\mathbf{B}| \cdot e^{-\frac{r-R_s}{\lambda}},
\end{equation}
where $H_0 = 4 \cdot 10^{-5}\;$erg cm$^{-3}$ s$^{-1}$ G$^{-1}$ and $\lambda=0.7\;R_\odot$.

{The last equation is introduced to ensure the divergence-free constrain $\nabla \cdot \vec{B} = 0$ and $\phi$ is a scalar potential function that constrains $\vec{B}$ to the space of divergence free fields. We use the artificial compressibility analogy \citep{chorin1997}, which is similar to hyperbolic divergence cleaning \citep{Dedner2002} and performs well with implicit solvers \cite{Yalim20211}. V$_{ref}$ denotes the propagation speed of the numerical divergence error. This way, the whole equation system remains hyperbolic.}

Due to high solar activity during the 2024 solar eclipse, we use a low-resolution mesh to avoid extreme gradients in the computational domain. We use an unstructured fifth-level subdivided geodesic polyhedron mesh starting from $1\;$R$_\odot$ to $30\;$R$_\odot$. The computational domain comprises approximately 380,000 finite volume elements. A detailed description of the used mesh is given by \cite{Brchnelova2022b}. We impose uniform initial boundary conditions for the plasma variables. Note that in this study, the computational domain starts above the transition region, and we use the characteristic parameter values for the base of the solar corona. We choose the commonly prescribed {values} for the density and temperature {characteristic for the base of the solar corona}, namely $n_\odot = 2\times 10^8\;$cm$^{-3}$ and $\text{T} = 1.8\times 10^{6}\;$K. We utilise the magnetic field derived from the magnetogram in a smoothing procedure to determine the inner boundary conditions for the radial magnetic field component. More details about the processing of the magnetogram are given in section~\ref{sec:input_magnetograms}. We use a Dirichlet boundary condition for the $B_r$ magnetic field component. The procedure for the magnetic field driving is similar to the one described in \cite{Kuzma2023} and \cite{Baratashvili2024C}. The magnetic field components in the other spatial directions (B$_\theta$ and B$_\phi$) are allowed to evolve freely and are defined as the corresponding values at the centroid of the nearest inner cell. The velocity is set so the plasma flows along the magnetic field lines \citep{Brchnelova2022}.

\section{Simulation setup} \label{sec:input_magnetograms}

\begin{figure*}[hpt!]
    \includegraphics[width=0.49\textwidth]{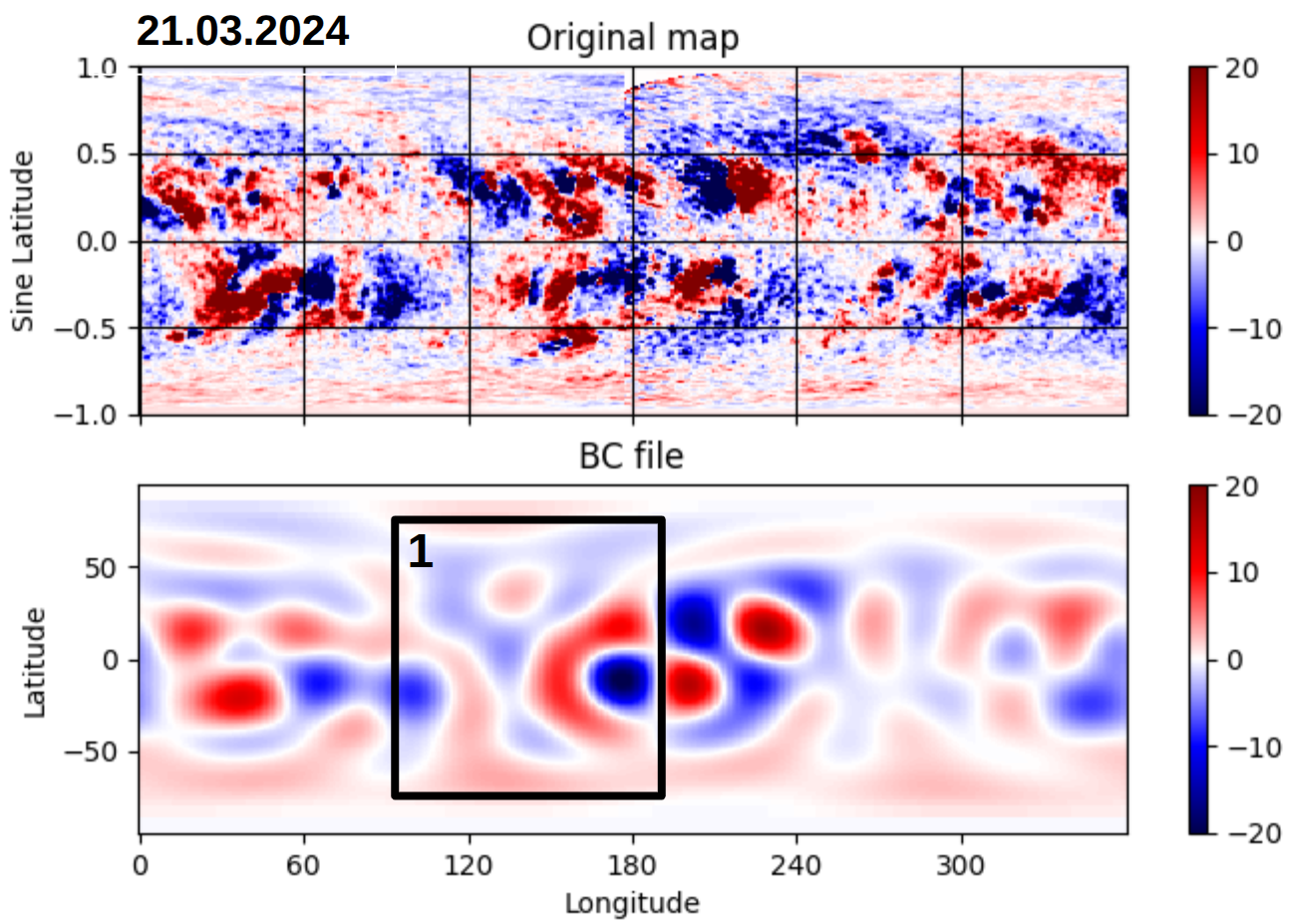}
    \hfill
    \includegraphics[width=0.49\textwidth]{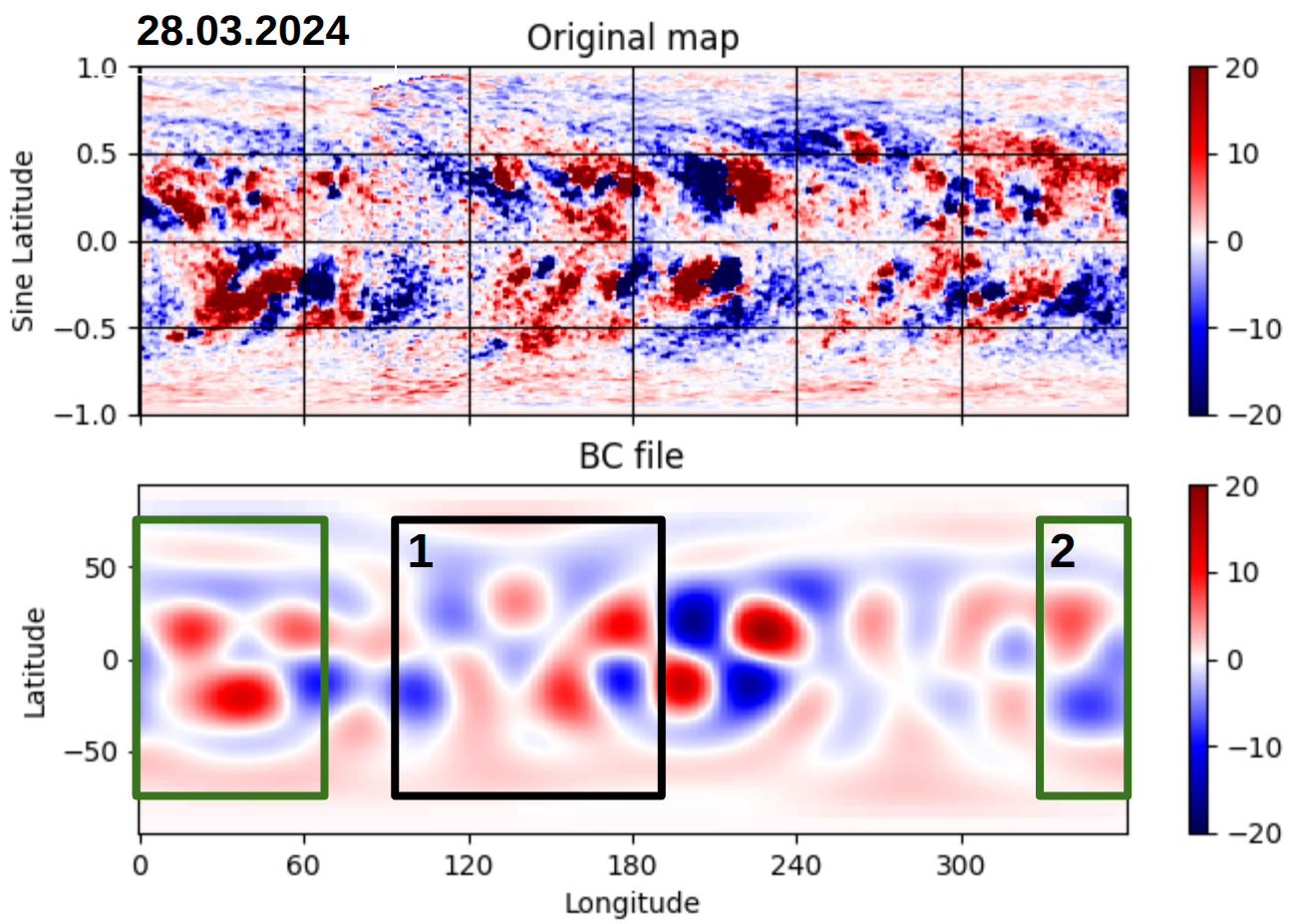}
    \hfill
    \includegraphics[width=0.49\textwidth]{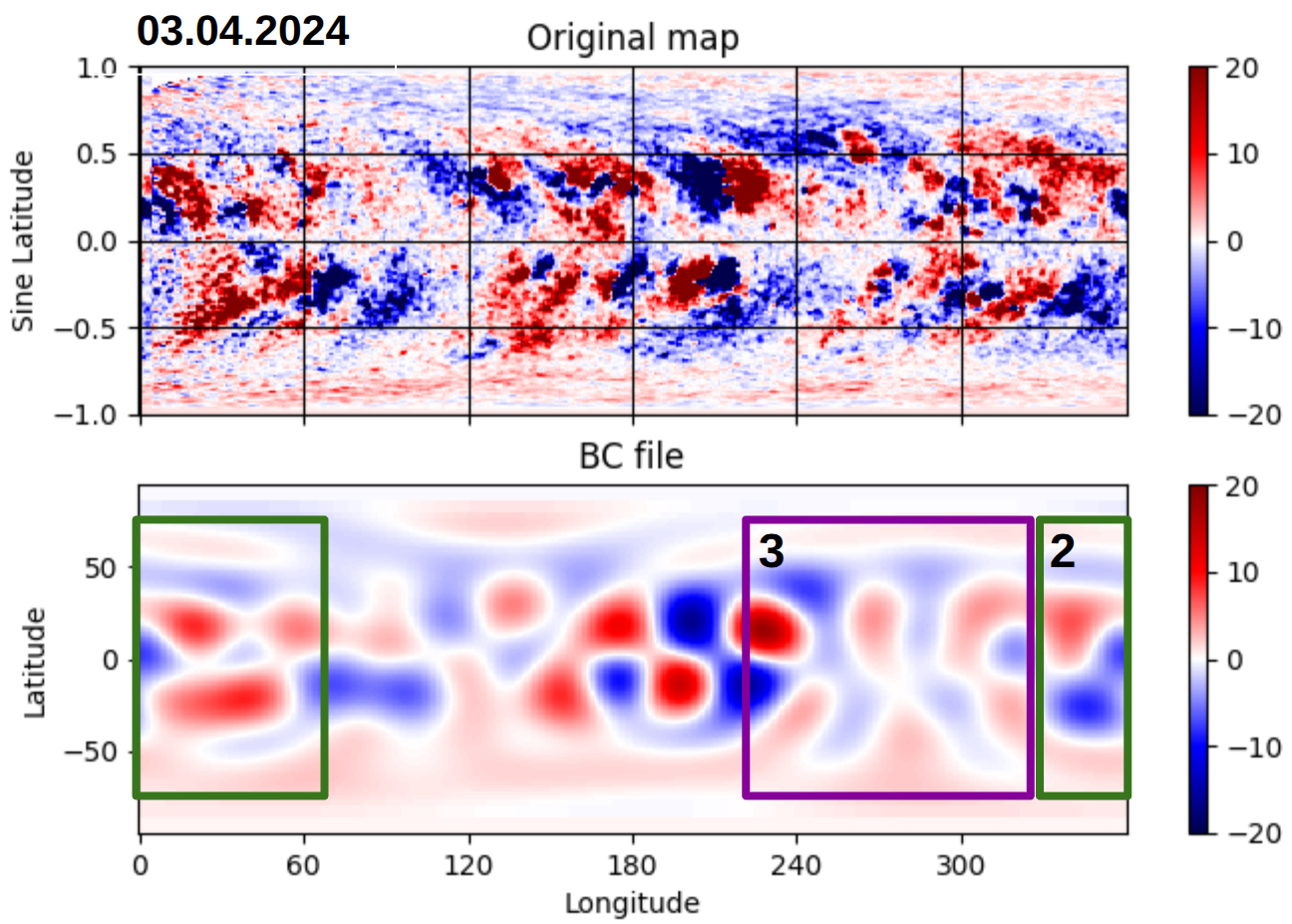}
    \hfill
    \includegraphics[width=0.49\textwidth]{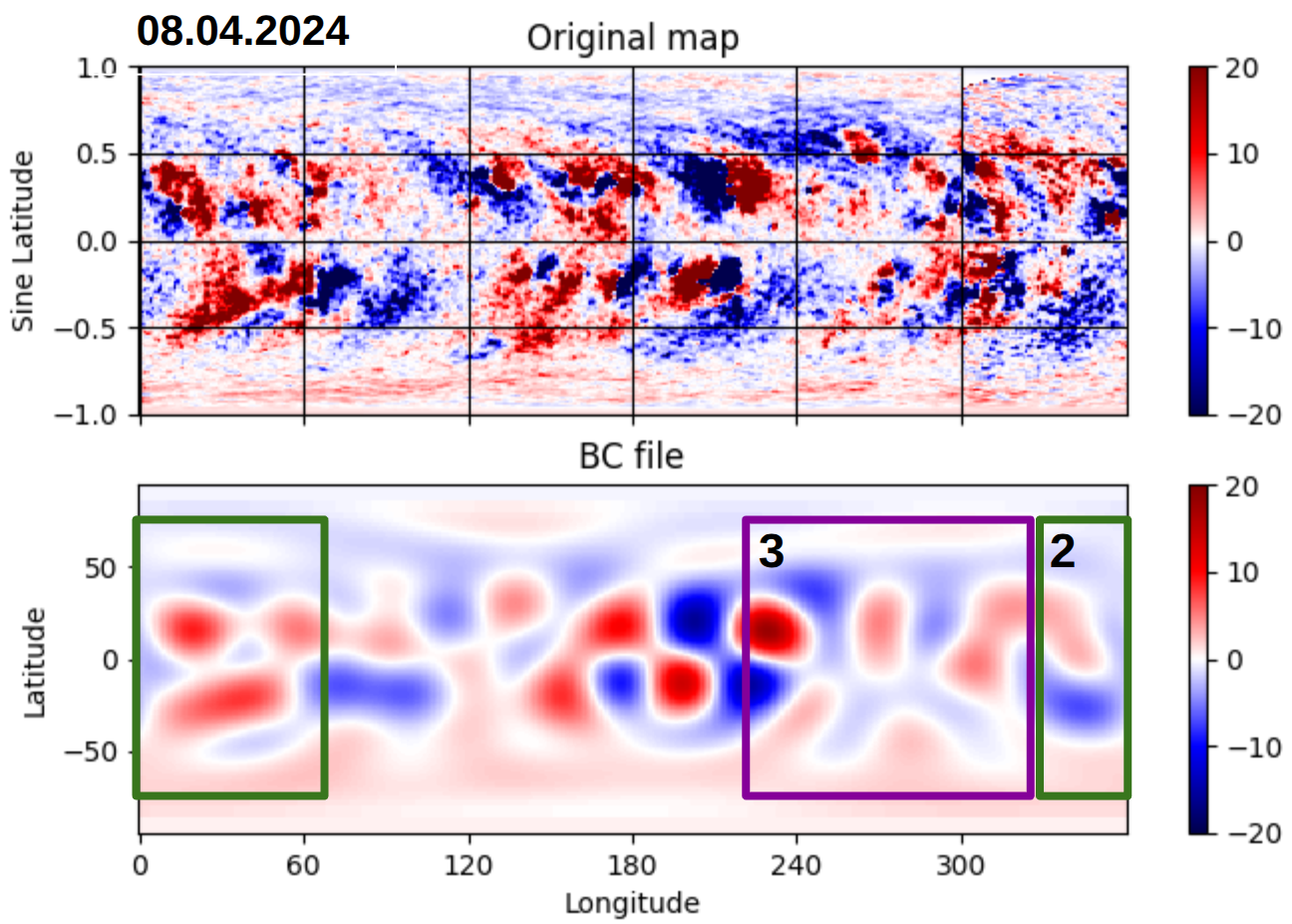}
    \hfill
  \caption{The original GONG (upper panels) and pre-processed input magnetograms (lower panels) for March 21, March 28, April 3 and April 8. The colour maps are saturated to [-20, 20]\;G range. All maps are rotated to align with the Earth's view angle as of April 8, 2024. Squares 1, 2 and 3 represent the areas of interest.}\label{fig:input_magnetograms}
\end{figure*}
The predictions for the total solar eclipse on April 8, 2024, started on March 21, 2024. We chose the magnetogram product from the GONG  (Global Oscillation Network Group) observatory. As reported in \cite{Perri2022}, \cite{Kuzma2023} and \cite{Baratashvili2024C}, a spherical harmonic decomposition was used to pre-process the maps. Due to a high solar activity, the spherical harmonics beyond $\ell_{max}=10$ were filtered out. Thus, the obtained maps are heavily smoothed. Initially, the predictions were done with steady simulations with a 24-hour cadence at 19:00 UT daily. This led to quasi-dynamic predictions of the total solar eclipse evolving over 18 days leading up to the eclipse day. We selected four points in time and corresponding snapshots from the entire prediction window to minimise excessive data and overwhelming information. 

Figure~\ref{fig:input_magnetograms} gives the original and processed magnetograms. The upper left magnetogram corresponds to the start of the prediction window on March 21. For better comparison, both original and processed magnetograms are saturated to the [-20, 20]\;G range for all shown magnetograms. The upper-right figure corresponds to the solar magnetic field configuration after 1 week. The rectangles indicate the regions where the most prominent difference between the plotted magnetograms was observed. These regions were identified both by looking at the structure of the magnetic field in the input map and the difference between the maps given in Figure~\ref{fig:Br_diff_images}. The rectangle labelled {`1'} is plotted on both magnetograms to show and emphasise that the magnetic field within this area changes significantly in this period. The difference is especially noticeable to the naked eye in the lower part of the rectangle. The lower-left figure corresponds to April 3. Here, a more substantial change can be spotted in the area enclosed in rectangle 2 compared to the March 28 solar magnetic field configuration. The lower-right figure corresponds to the day of the total solar eclipse. We can see that both regions enclosed by boxes 2 and 3 are significantly different from the magnetic field observed on April 3. These areas significantly affect the coronal configuration from the Earth's point of view, resulting in variable predictions as the eclipse day approaches. 

For the dynamic (time-dependent) regime, which is necessary to capture the continuous temporal evolution of the coronal structures, we set the time step to 5 minutes to balance computational efficiency with the desired numerical stability and accuracy \citep{Wang2025,wang2025sipifvmtimeevolvingcoronalmodel}. We drove the coronal evolution using both a series of hourly updated magnetograms and daily updated magnetograms. We applied a cubic Hermite interpolation to derive the inner boundary magnetic field at each physical time step \citep{Wang2025,wang2025sipifvmtimeevolvingcoronalmodel}.
For comparison with the quasi-steady predictions, the simulations with the time-evolving COCONUT version were performed with daily updated magnetograms and later with hourly updated magnetograms. In both cases, magnetograms at each physical time step were interpolated from the temporally nearest observed magnetograms. Thus, simulations in the quasi-steady and dynamic regimes were driven by the same magnetograms, with different cadences and driving modes.  

\section{Total solar eclipse predictions} \label{sec:predictions}
\begin{figure*}[hpt!]
\centering
    \includegraphics[width=0.49\textwidth]{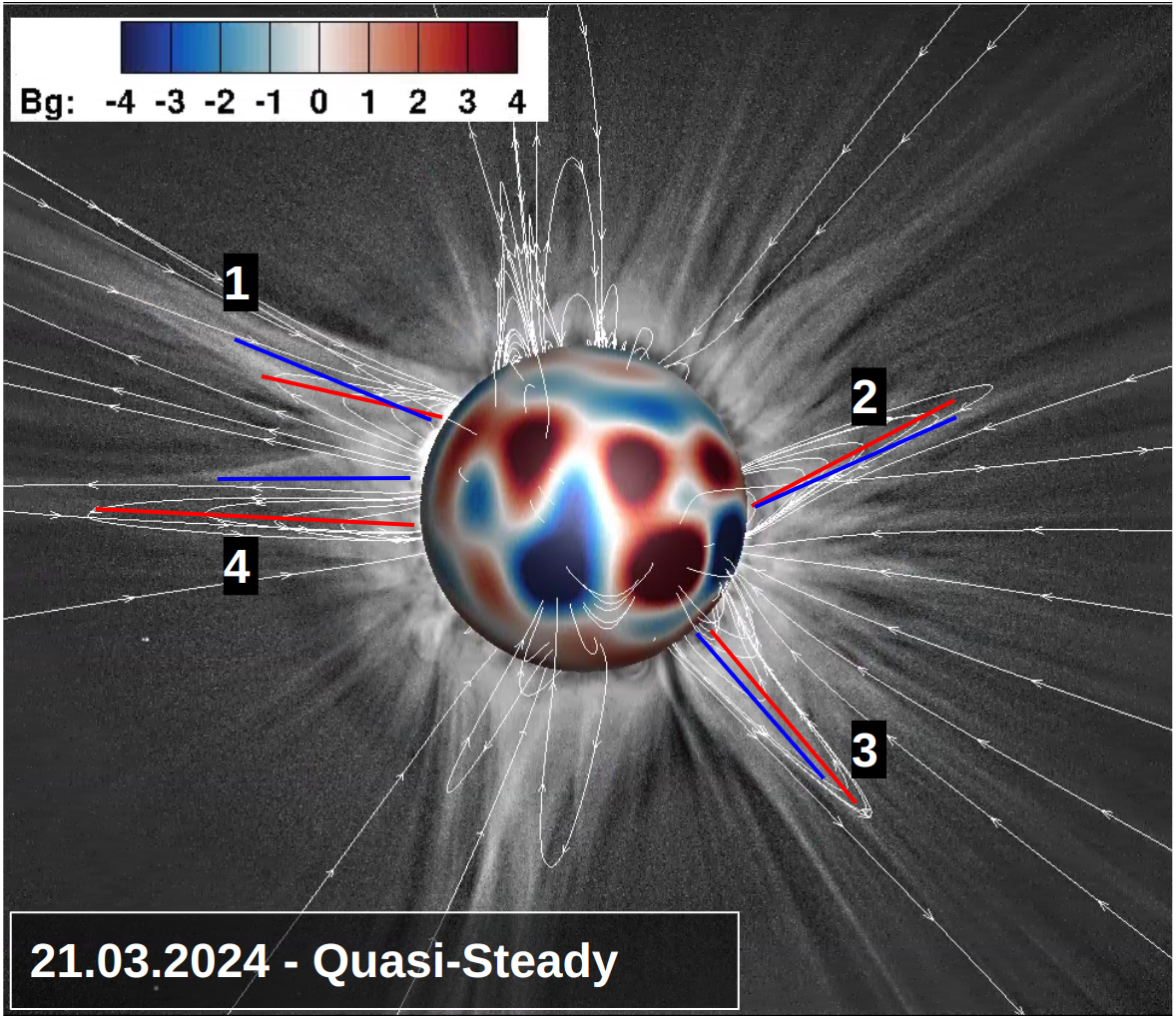}
    \hfill
    \includegraphics[width=0.49\textwidth]{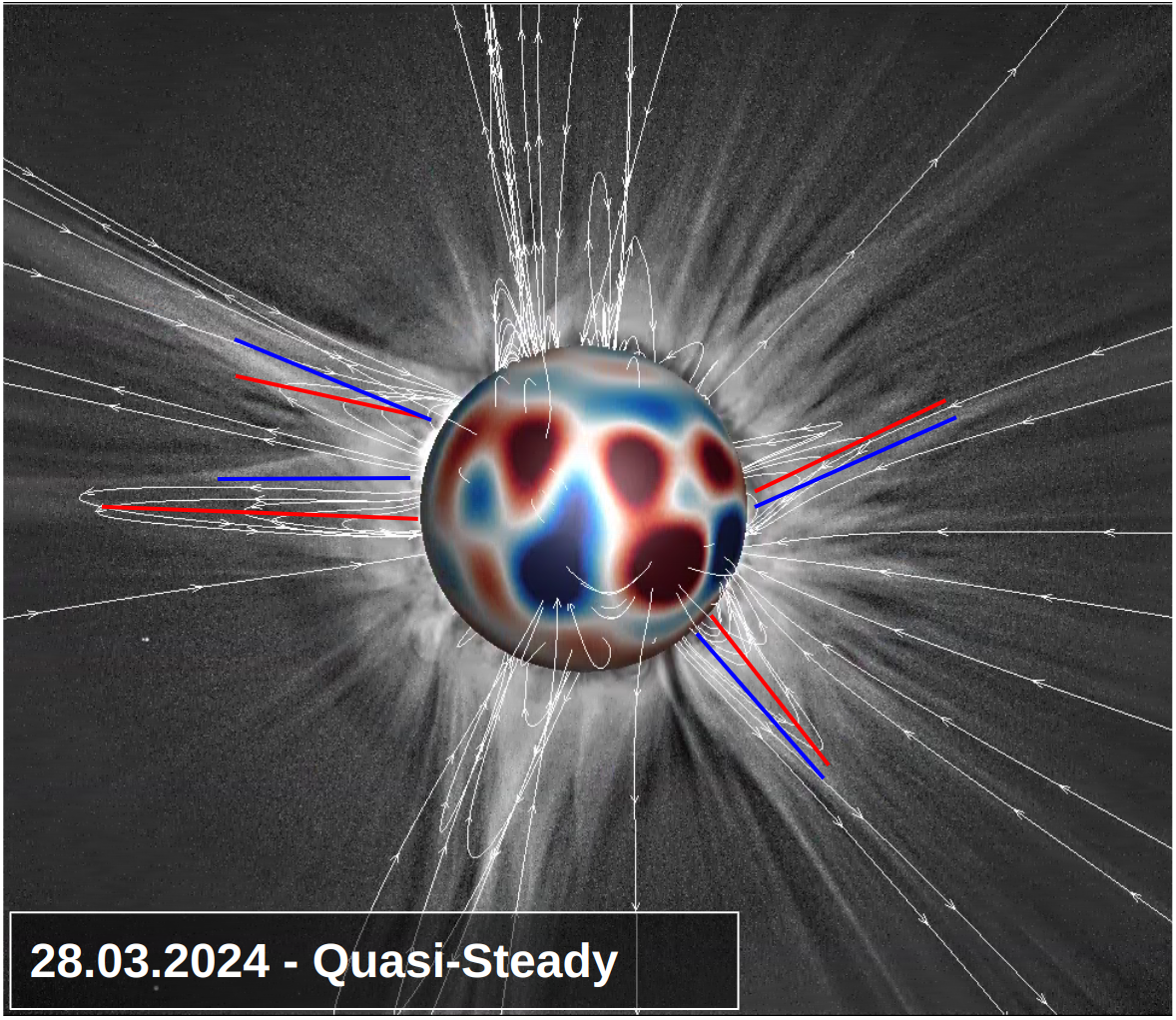}
    \hfill
    \includegraphics[width=0.49\textwidth]{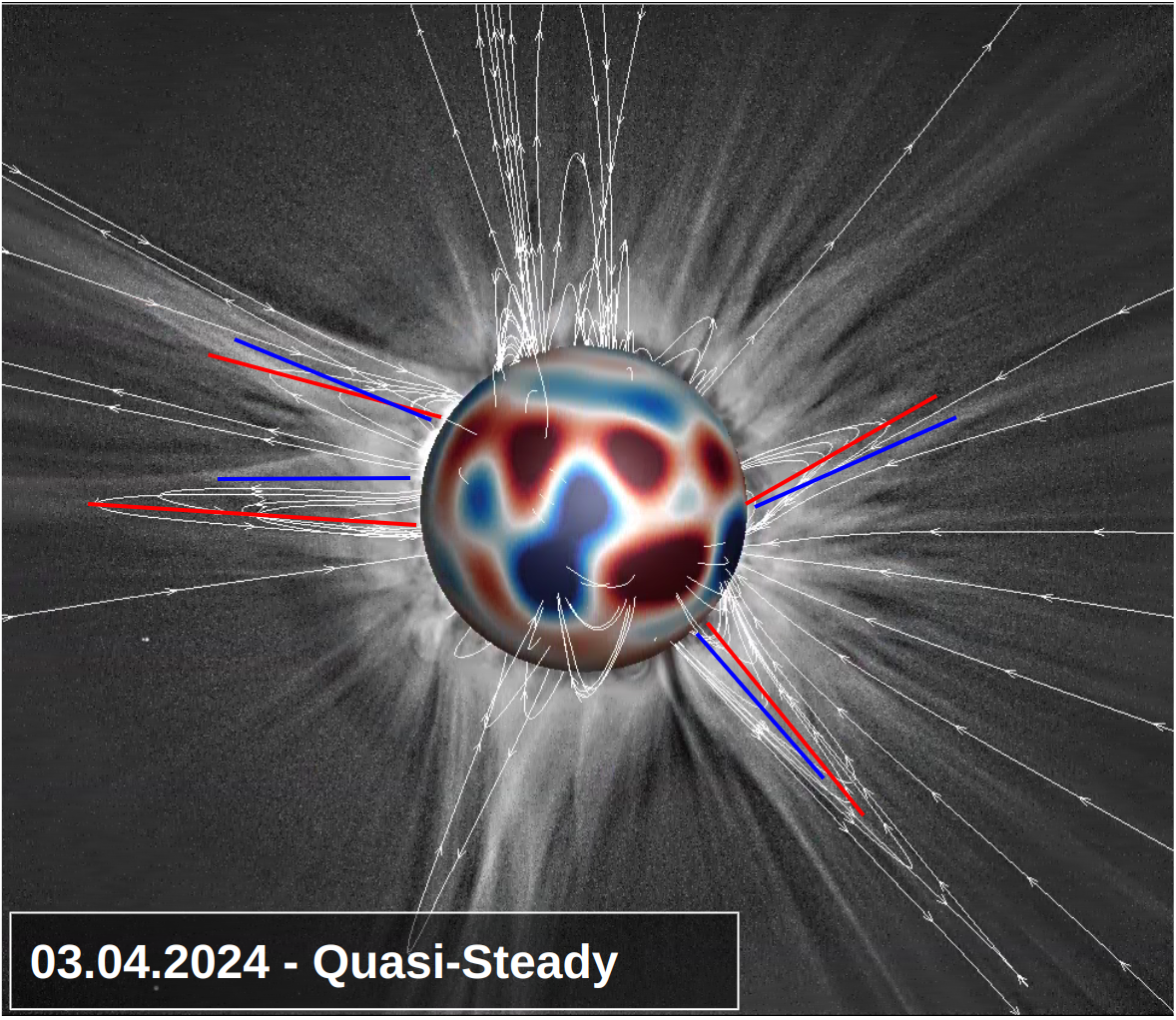}
    \hfill
    \includegraphics[width=0.49\textwidth]{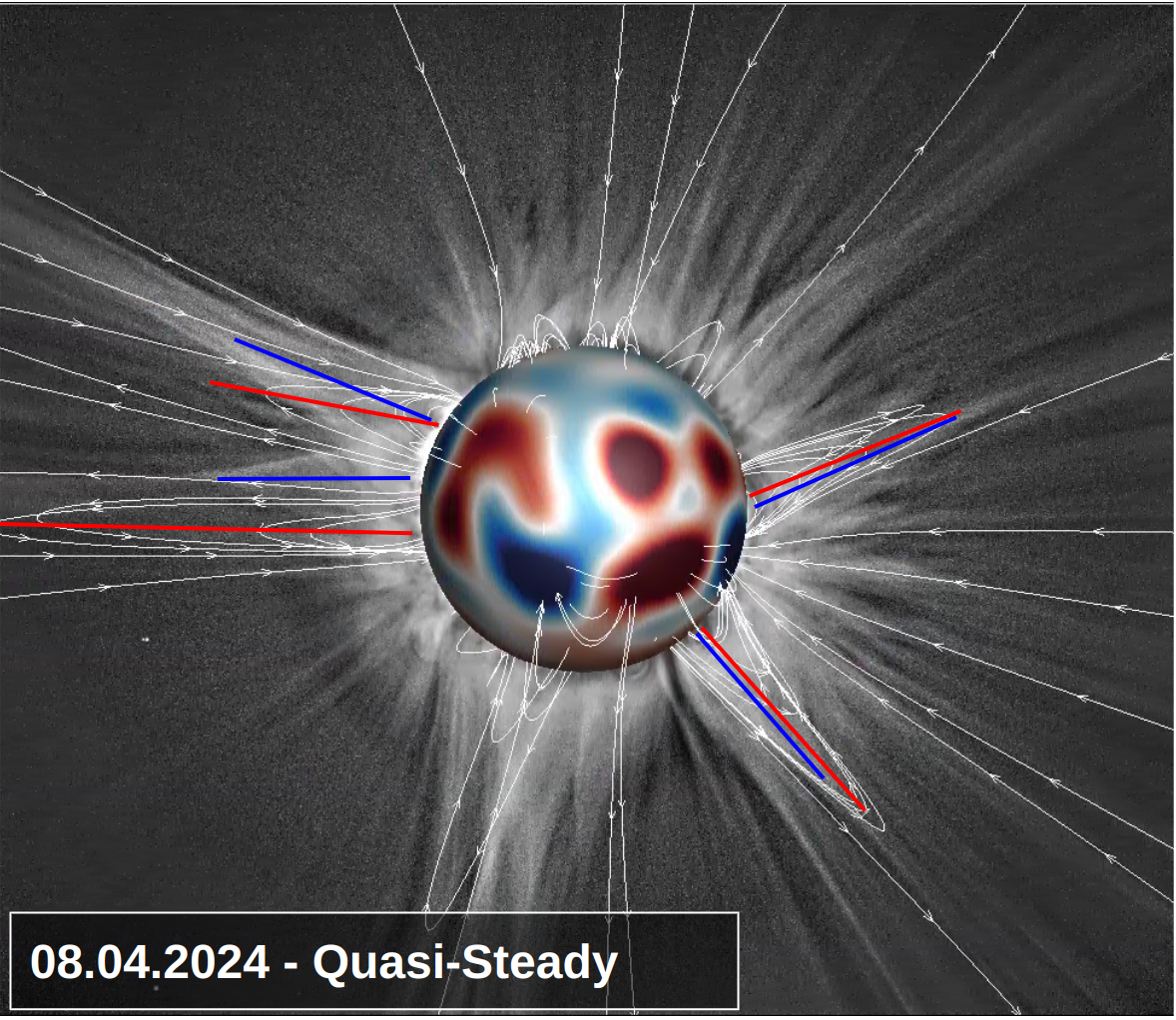}
  \caption{Total solar eclipse predictions by the COCONUT model in the quasi-steady regime. Each figure corresponds to the corona prediction for different magnetogram files from Figure~\ref{fig:input_magnetograms}. The solar surface is coloured with the radial magnetic field values saturated to [-4,4]\;G. The magnetic field lines are overplotted on the eclipse image. The blue lines indicate the directions of the selected streamers based on the eclipse image. Red lines correspond to the direction of the same streamers as modelled by COCONUT simulations. Eclipse image credits: Eclipse team of Nanjing University - Wu, Sizhe (photographer); Li, Yihua; Huang, Yuhao; Lao, Qinghui; Cheng, Xin; Qu, Zhongquan.} \label{fig:quasi_steady_streamerss}
\end{figure*}
\begin{figure*}[hpt!]
\centering
    \includegraphics[width=0.49\textwidth]{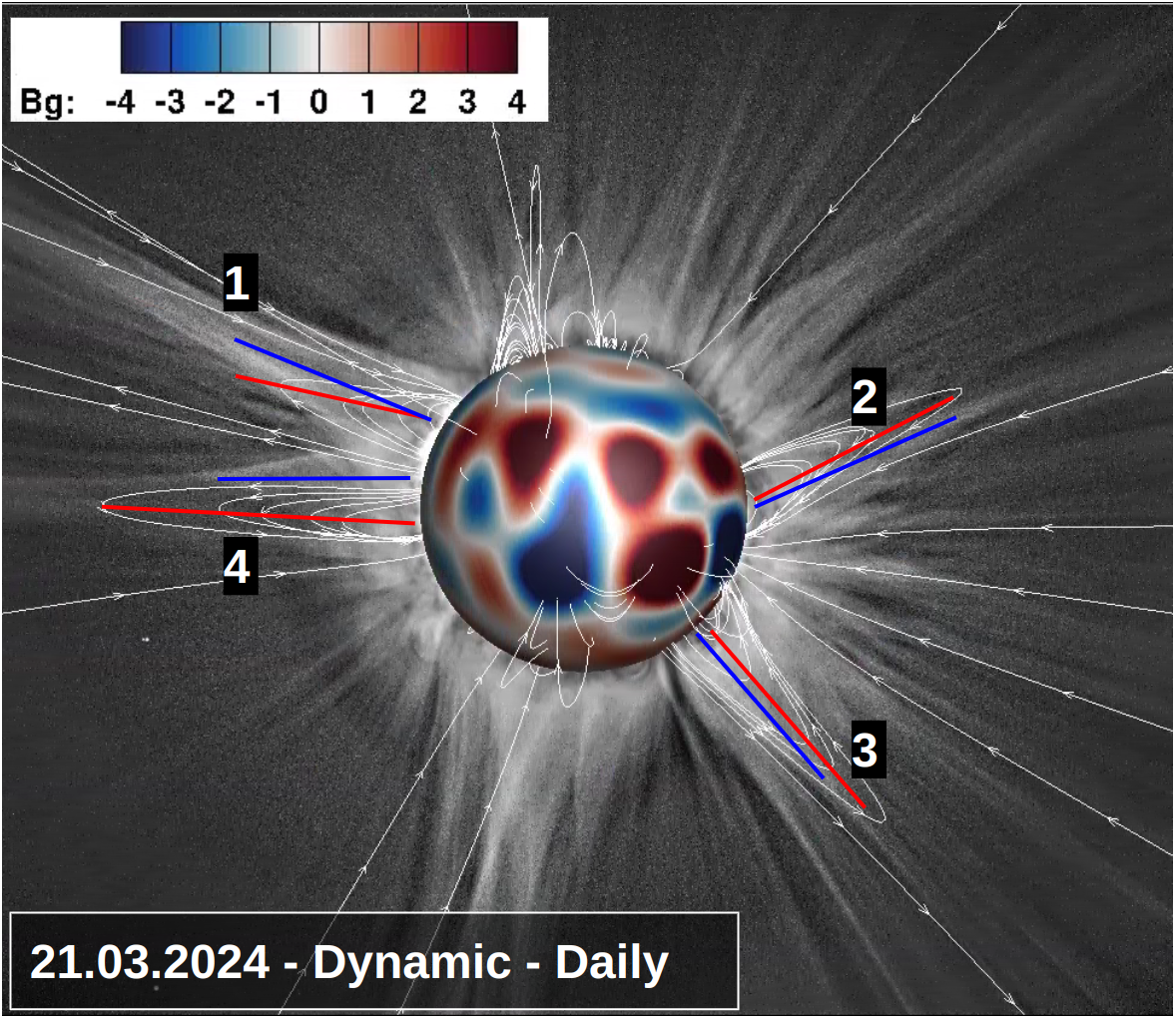}
    \hfill
    \includegraphics[width=0.49\textwidth]{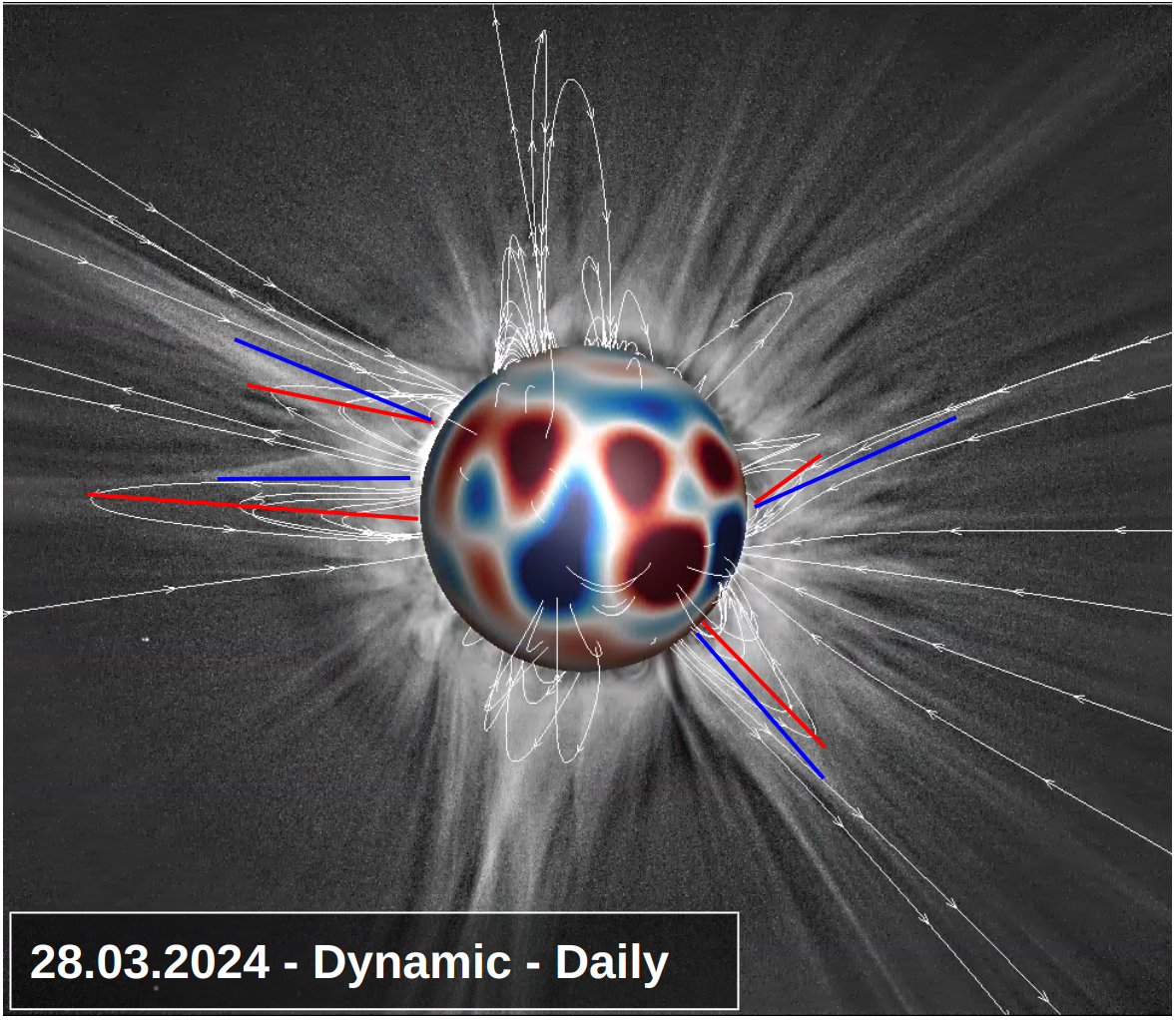}
    \hfill
    \includegraphics[width=0.49\textwidth]{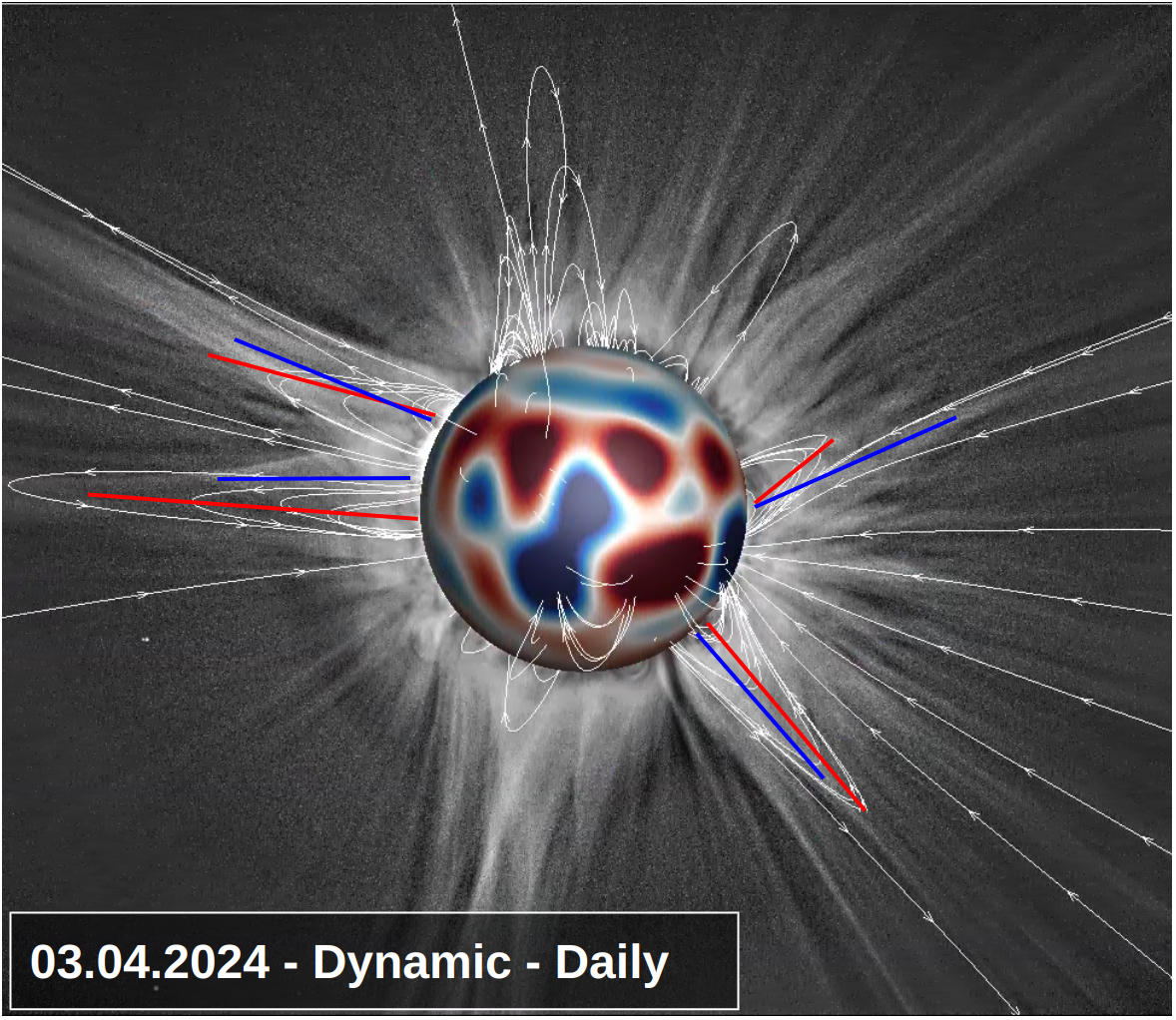}
    \hfill
    \includegraphics[width=0.49\textwidth]{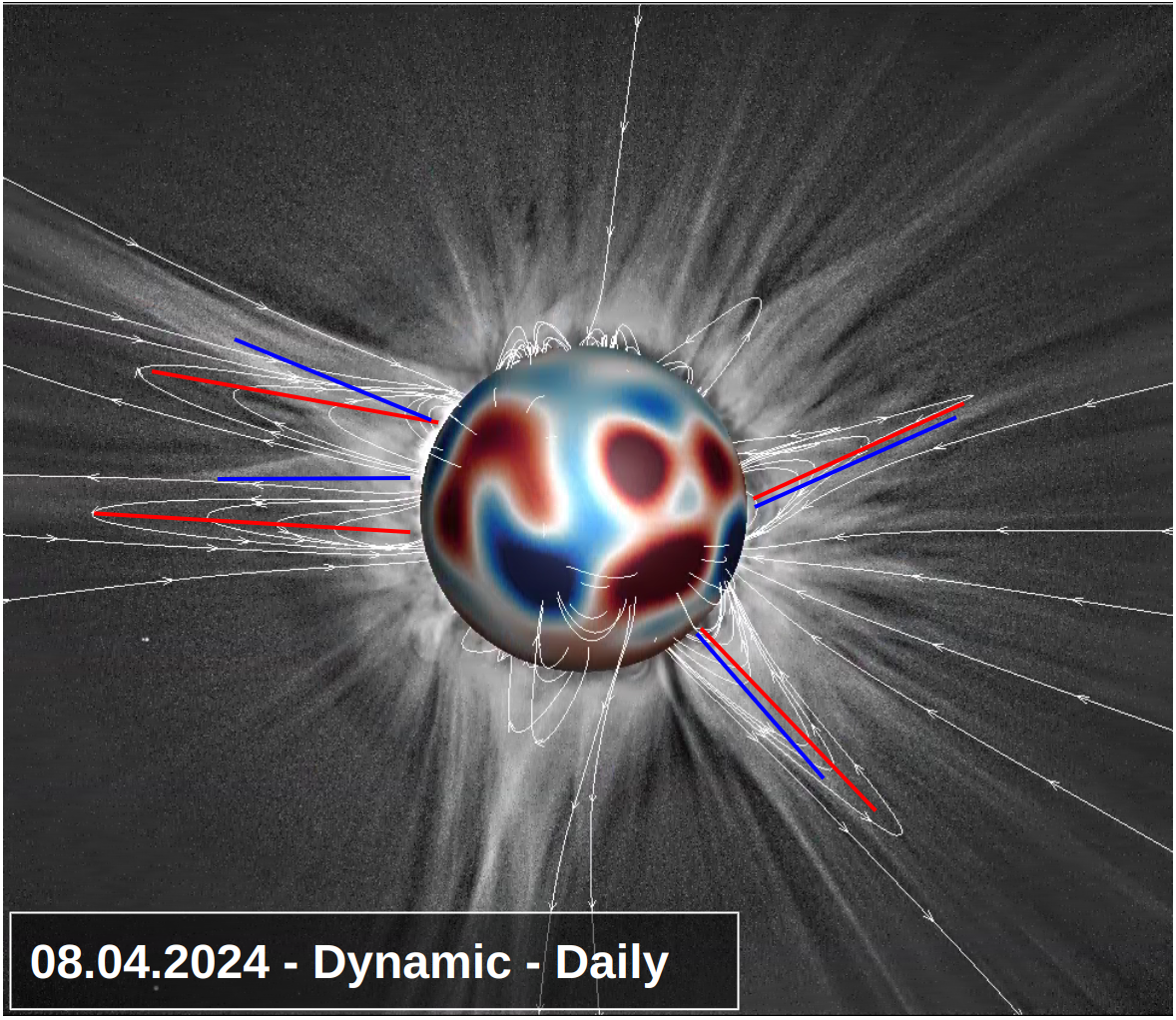}
  \caption{Total solar eclipse predictions by the COCONUT model in the dynamic regime with daily updated magnetograms. Each figure corresponds to the corona prediction for different magnetogram files from Figure~\ref{fig:input_magnetograms}. The solar surface is colored with the radial magnetic field values saturated to [-4,4]\;G. The magnetic field lines are overplotted on the eclipse image. The blue lines indicate the directions of the selected streamers based on the eclipse image. Red lines correspond to the direction of the same streamers as modelled by COCONUT simulations. Eclipse image credits: Eclipse team of Nanjing University - Wu, Sizhe (photographer); Li, Yihua; Huang, Yuhao; Lao, Qinghui; Cheng, Xin; Qu, Zhongquan. }\label{fig:dynamic_daily_streamers}
\end{figure*}

\begin{figure*}[hpt!]
\centering
    \includegraphics[width=0.49\textwidth]{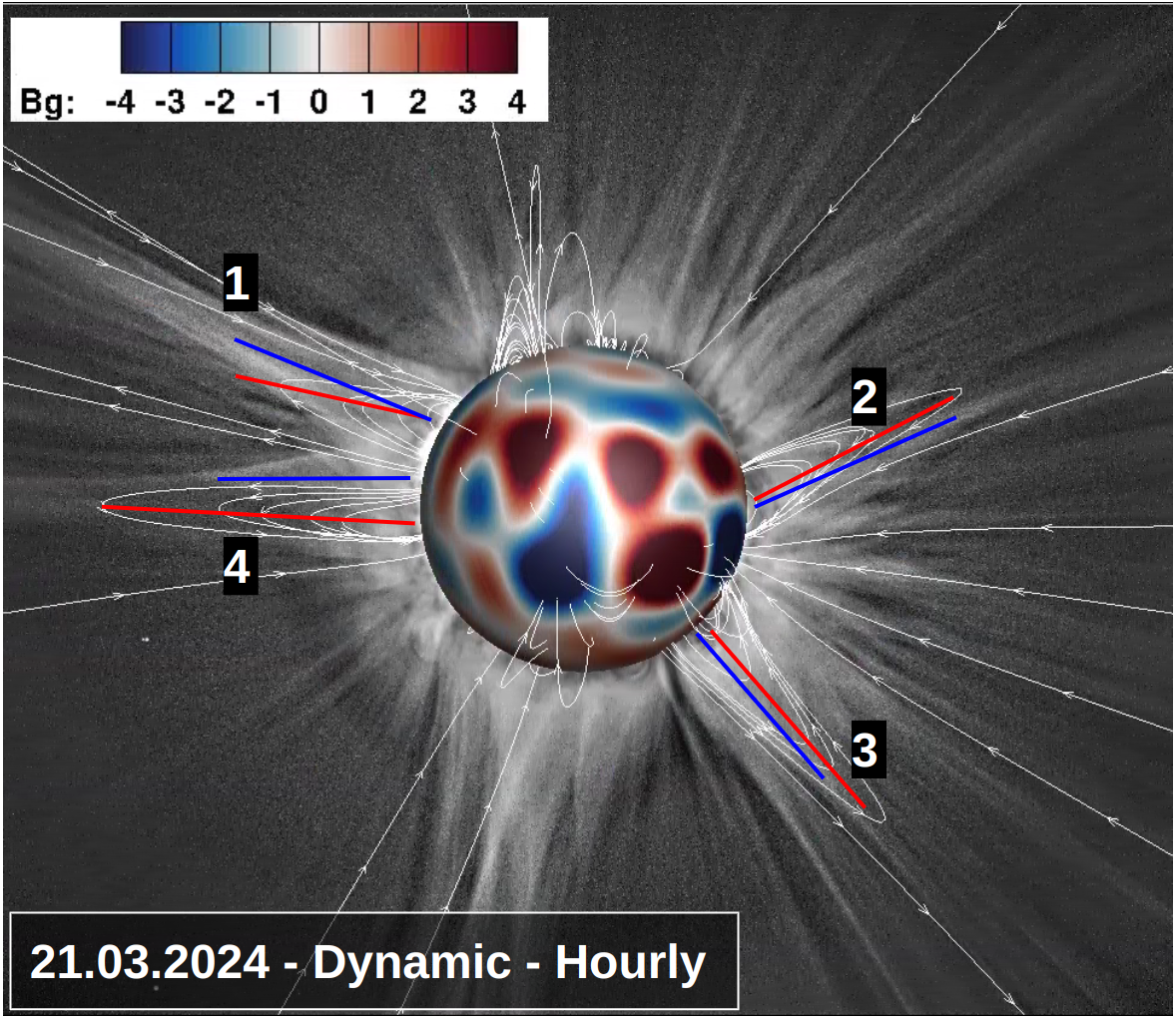}
    \hfill
    \includegraphics[width=0.49\textwidth]{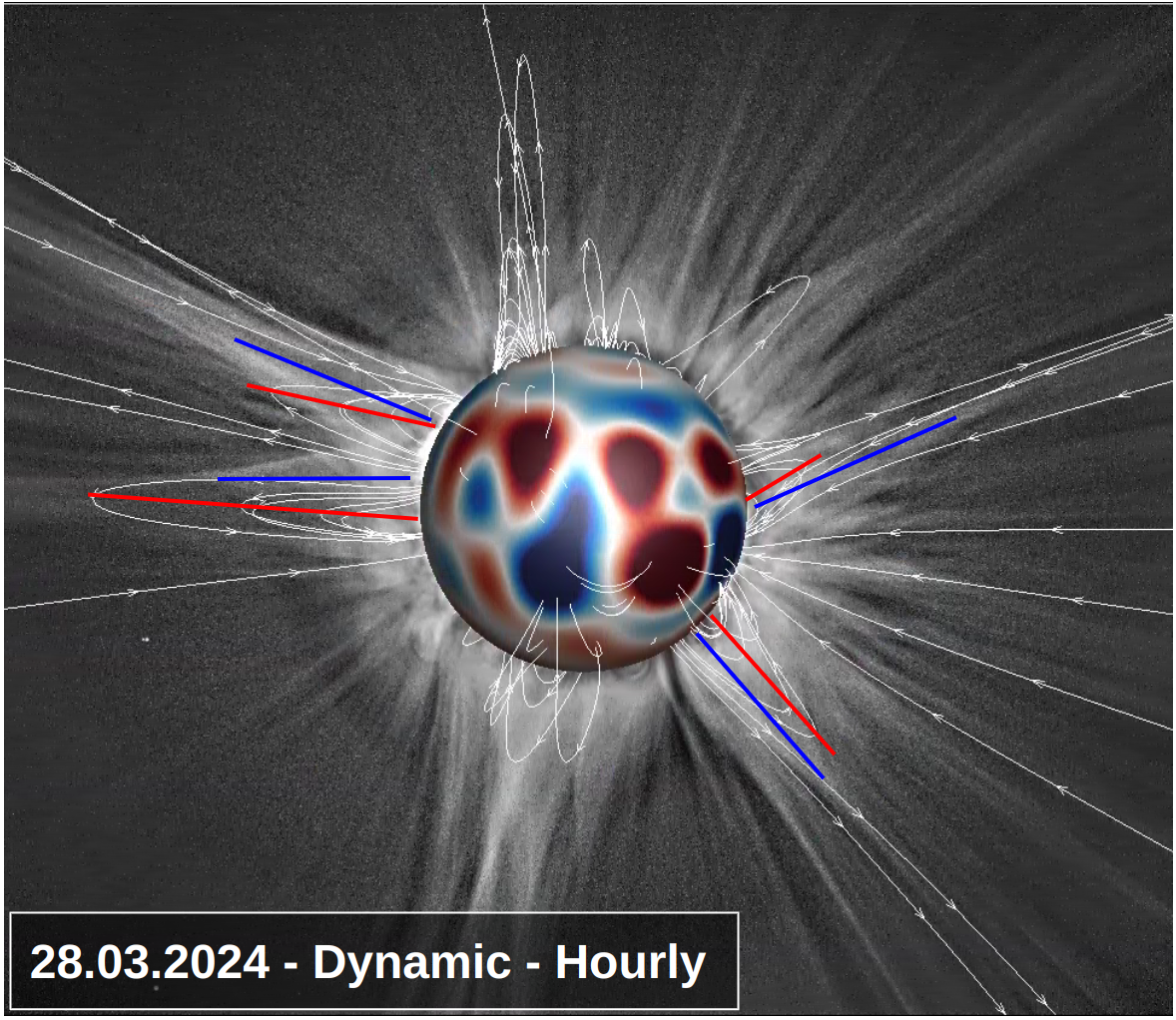}
    \hfill
    \includegraphics[width=0.49\textwidth]{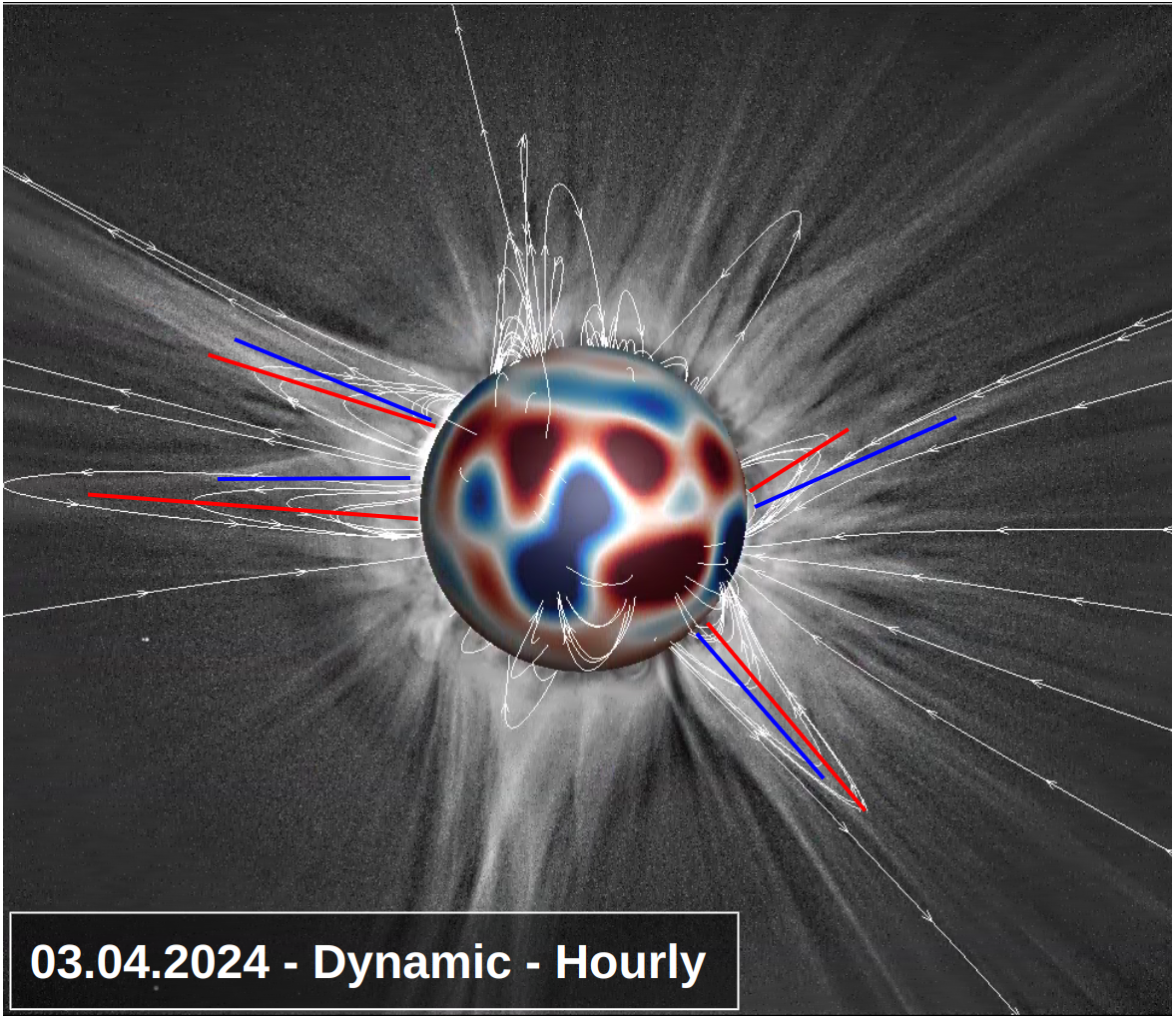}
    \hfill
    \includegraphics[width=0.49\textwidth]{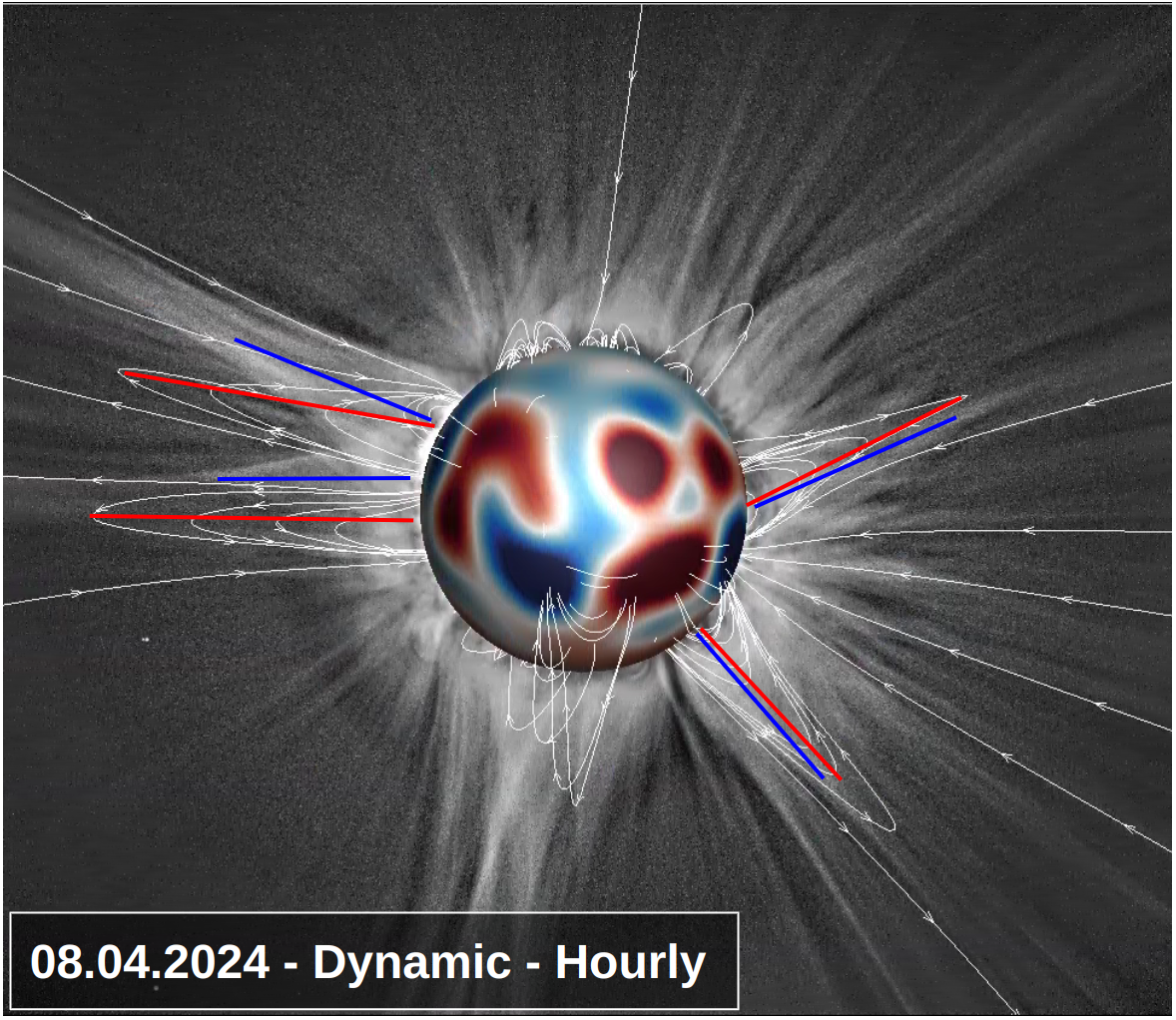}
  \caption{Total solar eclipse predictions by the COCONUT model in the dynamic regime with hourly updated magnetograms. Each figure corresponds to the corona prediction for different magnetogram files from Figure~\ref{fig:input_magnetograms}. The solar surface is colored with the radial magnetic field values saturated to [-4,4]\;G. The magnetic field lines are overplotted on the eclipse image. The blue lines indicate the directions of the selected streamers based on the eclipse image. Red lines correspond to the direction of the same streamers as modelled by COCONUT simulations. Eclipse image credits: Eclipse team of Nanjing University - Wu, Sizhe (photographer); Li, Yihua; Huang, Yuhao; Lao, Qinghui; Cheng, Xin; Qu, Zhongquan.}\label{fig:dynamic_houry_streamers}
\end{figure*}

The total solar eclipse predictions were made in quasi-steady and dynamic regimes. The effect of updating the inner boundary conditions was also investigated. This study aims to assess which approach yields more accurate predictions. Since the input magnetic field was smoothened and a low-resolution computational grid was used, we used various techniques to validate the modelled corona and present the most comprehensive assessment of its strengths and weaknesses. Below, all different regimes are considered separately, first with the coronal validation method reported by \cite{Wagner2022} {section 3.2.1 in the cited paper. The model focuses on coronal streamers, which are shaped by the Sun's magnetic field structure. \cite{Wagner2022} utilises the streamer directions in the Schatten Current Sheet (SCS) \citep{Schatten1969} model for the comparison with the coronagraph images from SOHO/LASCO-C2 or Solar Terrestrial Relations Observatory (STEREO) A - Cor2 images in the outer corona. In our case, we identify the loops and compare them to the direction of the total solar eclipse, as the image also includes the lower corona.} The same seeds were used for the magnetic field lines for all simulation results, so the comparison is as objective as possible.  

\subsection{Quasi-steady simulations}

The predictions reported in this section are made daily from March 21, 2024, onwards. Each simulation is a steady-state simulation driven by the generated magnetic map based on the GONG magnetograms described in Section~\ref {sec:input_magnetograms}. The modelled coronas are rotated such that zero longitude corresponds to April 8, the day of the total eclipse, from each input magnetogram. Each of these simulations takes $~\sim20$ minutes on four nodes with 36 cores each, on the Genius cluster of the Flemish Supercomputer centre (VSC)\footnote{\url{https://www.vscentrum.be/}}. Figure~\ref{fig:quasi_steady_streamerss} shows the total solar eclipse image taken by the Eclipse team of Nanjing University overlaid with the modelled magnetic field lines in COCONUT. The processed radial magnetic field used for the modelling is plotted on the solar surface. The magnetic field is saturated to [-4, 4]~G to emphasise the sharper features. The upper left figure corresponds to the simulation driven by the magnetogram observed on March 21, and the upper right figure shows the modelled corona for the March 28 magnetogram. The bottom left and right figures correspond to total solar eclipse predictions driven by April 3 and April 8 magnetograms, respectively. The blue and red lines denote the streamer directions on each figure, estimated similarly as demonstrated by \cite{Wagner2022}. Since the total eclipse occurred during the maximum activity phase of the Sun, many streamers and loop signatures are present close to the solar surface. We chose to consider only four regions for this study, as including all of them would introduce a high probability of overlap and inaccurate estimations. We distinguished the four most prominent signatures and focused on these to compare them with this method. Blue lines correspond to the directions determined in the taken image; therefore, they are identical in each figure. Red lines correspond to the streamer directions identified from the COCONUT simulation results. 

On the east limb, we chose the regions corresponding to 1 and 4 in the figure, while on the west limb, regions 2 and 3 showed the strongest features. The signatures near the poles were difficult to distinguish and poorly reconstructed by the magnetic field lines in the COCONUT simulations. Therefore, we decided not to focus on these areas, as they would not allow us to assess the performance of the COCONUT model. 
First, let us consider the steamers on the east limb. We can see that steamer 1 in COCONUT is modelled closely to the observed one, but the streamer direction is different. For streamer 4, on the other hand, the observed and simulated streamer directions are parallel, but the locations are not close to each other. On the west limb, we can see that the regions corresponding to 2 and 3 are modelled closely to the observations. We do not observe a significant change in the simulation results on March 28 because the eclipse date is still 11 days away, and the new information provided by the updated magnetogram is not yet accurate enough to improve the prediction for April 8. On April 3, the magnetic field configuration differed slightly in the centre of the disk and the eastern part. The resulting streamers are slightly more aligned with the observed structures than in the previous case. On the eclipse day, the magnetic field appears quite different across the entire disk. This means that the solar surface magnetic field evolved significantly over the 5 days between April 3 and April 8. We can see that the west limb is more accurate regarding the location and direction of the streamers. However, on the east limb, there is no significant improvement. 

\subsection{Dynamic simulations}
\subsubsection{Daily update cadence}

The total solar eclipse predictions presented in this section are performed using the time-evolving COCONUT regime. In this regime, the input magnetograms are updated to yield time-dependent inner boundary conditions. First, we update the magnetograms daily to obtain inner boundary information similar to that in the quasi-steady regime. In this case, we obtain a continuously updated corona within the prediction window; however, we only plot the snapshots corresponding to the same days as in the quasi-steady case for comparison purposes. Figure~\ref{fig:dynamic_daily_streamers} shows the modelled corona. The figure is arranged similarly to figure~\ref{fig:quasi_steady_streamerss}. The eclipse image is overlaid by magnetic field lines from time-evolving COCONUT simulations. The first image, corresponding to the predictions driven by the magnetogram observed on March 21, is similar to the image from the quasi-steady regime. On March 28 and April 3, the steamer region denoted by number 2 is less aligned with the one in the eclipse image than that on March 21. All other streamers remain oriented in a manner similar to the previous date. The direction of this streamer also differs from that of the other dates in the quasi-steady regime. On April 8, we can see that both simulated streamers on the west limb are well aligned with the observed streamers. However, the two streamers on the east limb seem to come closer to each other. Streamer 4 is parallel to the direction of the corresponding streamer in the eclipse image. In contrast, the direction of streamer 1 appears to be less in agreement with the one observed on April 3. 
On the west limb, the quasi-steady and dynamic modelling regimes differ, while on the east limb, both regimes perform poorly. The two simulated streamers are in the right places, but their direction seems different. Note that, due to the maximum solar activity, distinguishing between different loops and streamers is extremely difficult, which could lead to errors when estimating the directions of the streamers in the eclipse images. 

\begin{figure*}[hpt!]
\centering
    \includegraphics[width=0.33\textwidth]{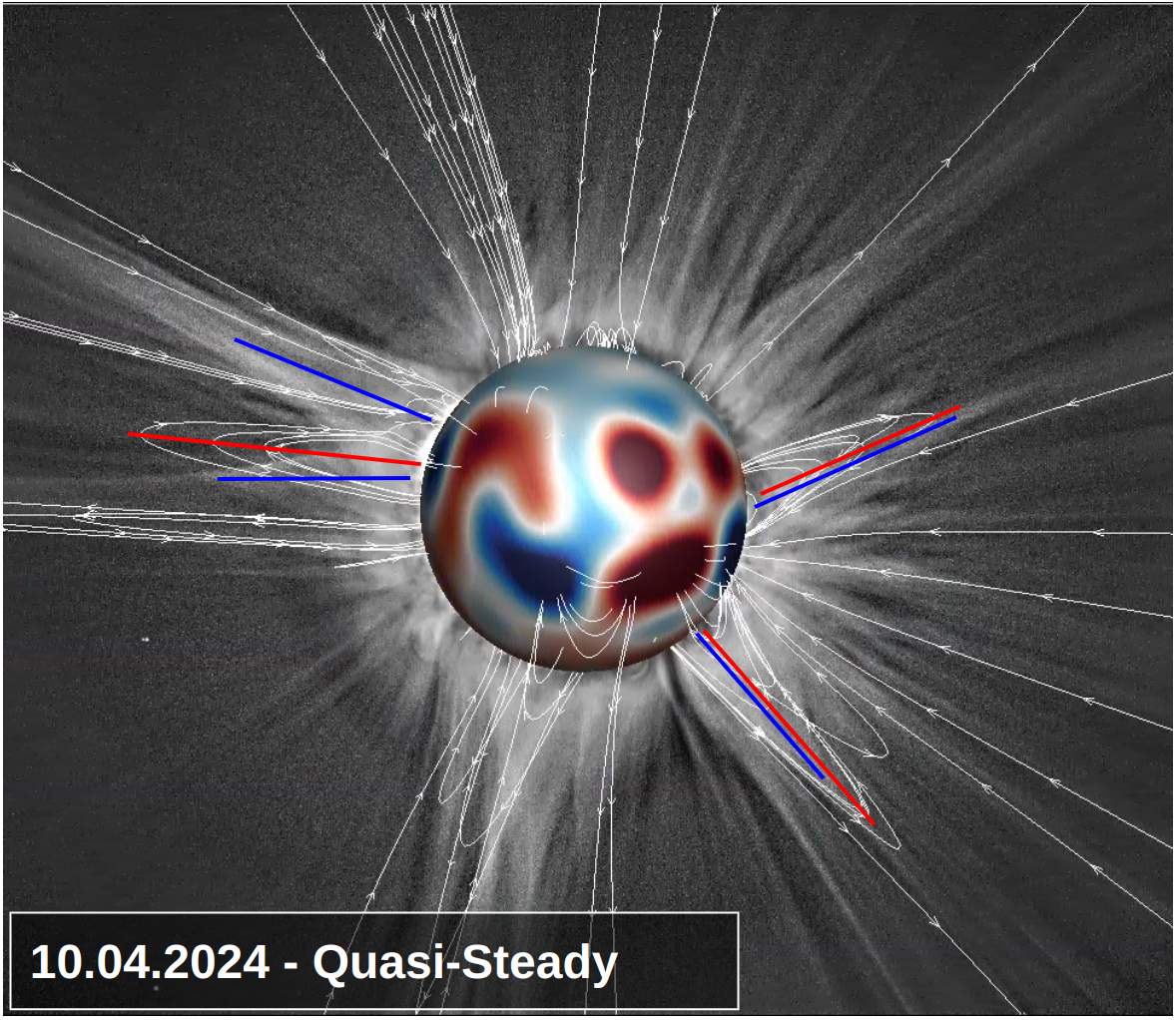}
    \hfill
    \includegraphics[width=0.33\textwidth]{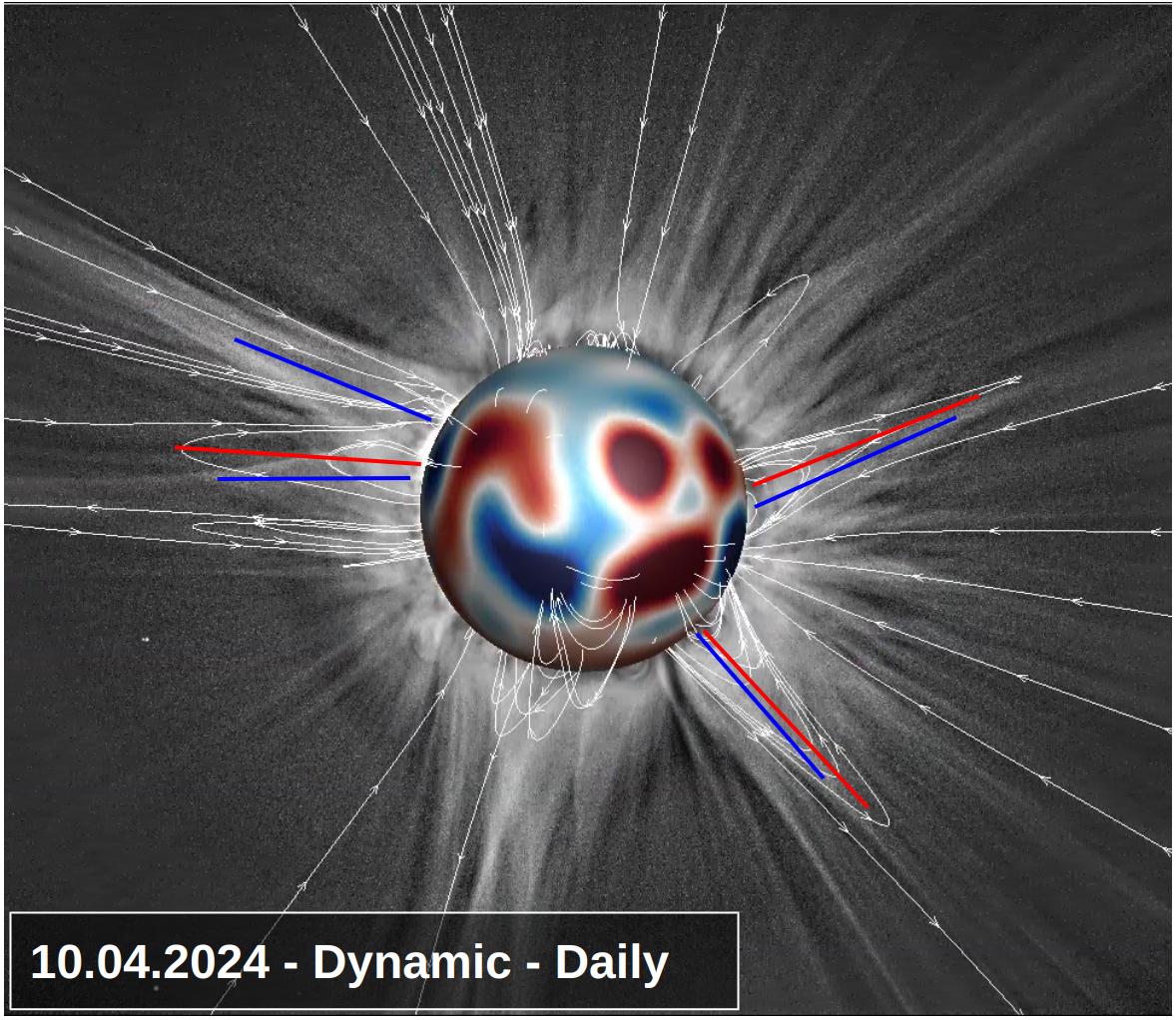}
    \hfill
    \includegraphics[width=0.33\textwidth]{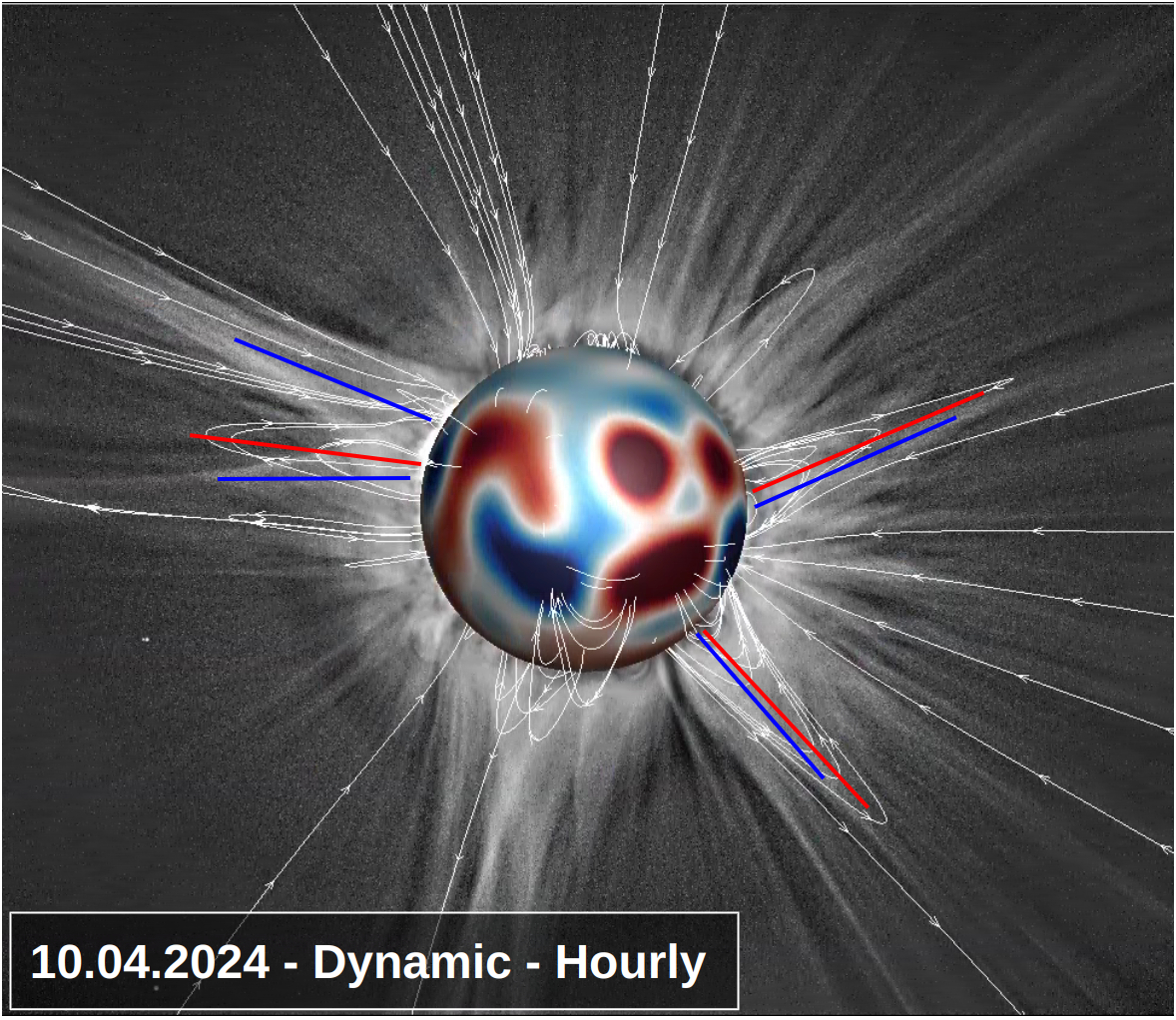}
  \caption{Total solar eclipse predictions by the COCONUT model with the magnetogram observed on April 10 in the quasi-steady, dynamic daily cadence and dynamic hourly regimes. The solar surface is colored with the radial magnetic field values saturated to [-4,4]\;G. The magnetic field lines are overplotted on the eclipse image. The blue lines indicate the directions of the selected streamers based on the eclipse image. Red lines correspond to the direction of the same streamers as modelled by COCONUT simulations. Eclipse image credits: Eclipse team of Nanjing University - Wu, Sizhe (photographer); Li, Yihua; Huang, Yuhao; Lao, Qinghui, Cheng, Xin; Qu, Zhongquan.}\label{fig:10Apr_steady_dynamic}
\end{figure*}

\subsubsection{Hourly update cadence}

This section presents results from the time-evolving COCONUT simulations, which were obtained by updating the inner boundary conditions at an hourly cadence.  Hence, 601 magnetograms were retrieved and pre-processed for this simulation.
This simulation was performed using 270 CPUs on the big-memory nodes of the WICE cluster at VSC, with each CPU allocated 2000\;MB of memory storage. It ran approximately 60 times faster than the real-time coronal evolution, completing a 600-hour physical period by just about 10 hours of computational wall-clock time.

Like the previous two cases, Figure~\ref{fig:dynamic_houry_streamers} shows the results for the modelled total solar eclipse coronal configuration corresponding to March 21, March 28, April 3 and April 8. When we compare the results in this case to those in the dynamic simulations with updated inner boundary conditions at a daily cadence, we can see that the results are similar, with only minimal differences. This is somewhat surprising because, considering the maximum solar activity, one would expect different coronal configurations with a higher updating frequency of the inner boundary magnetic field. When the photospheric magnetic field is updated hourly, all the available observational information is used to drive the dynamic solar corona. The magnetic field evolves rapidly on the solar surface near solar maximum. However, because of the applied pre-processing of the magnetic field and the obtained smoothed magnetic maps, the changes in the modelled total solar corona are not so significant when using the magnetograms with 24-hour cadence or 1-hour cadence. As a result, hourly dynamic simulations yield results that are not significantly different from those of daily simulations. 

\subsubsection{Post eclipse analysis}
\begin{figure*}[hpt!]
\centering
    \includegraphics[width=0.33\textwidth]{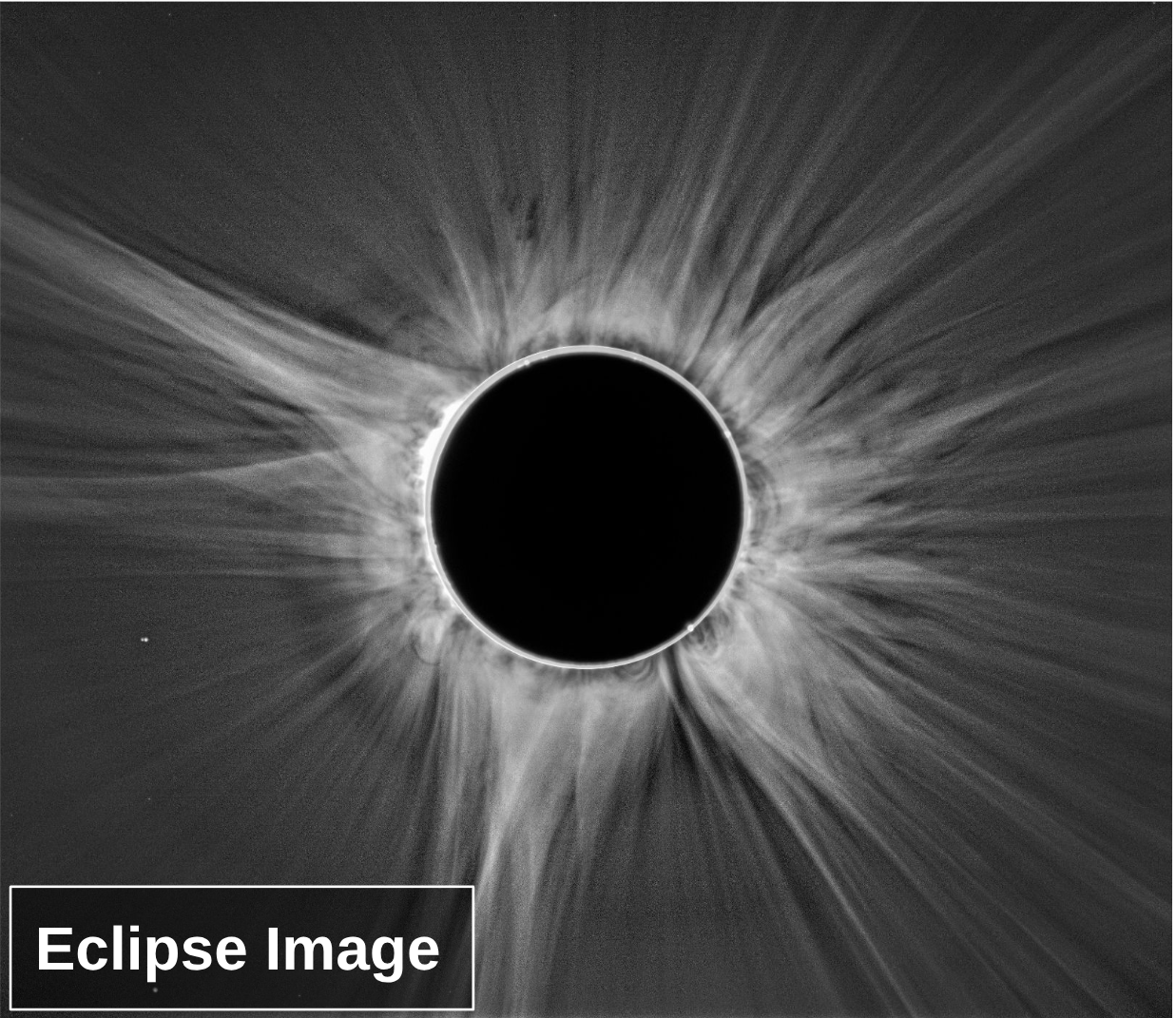}
    \includegraphics[width=0.33\textwidth]{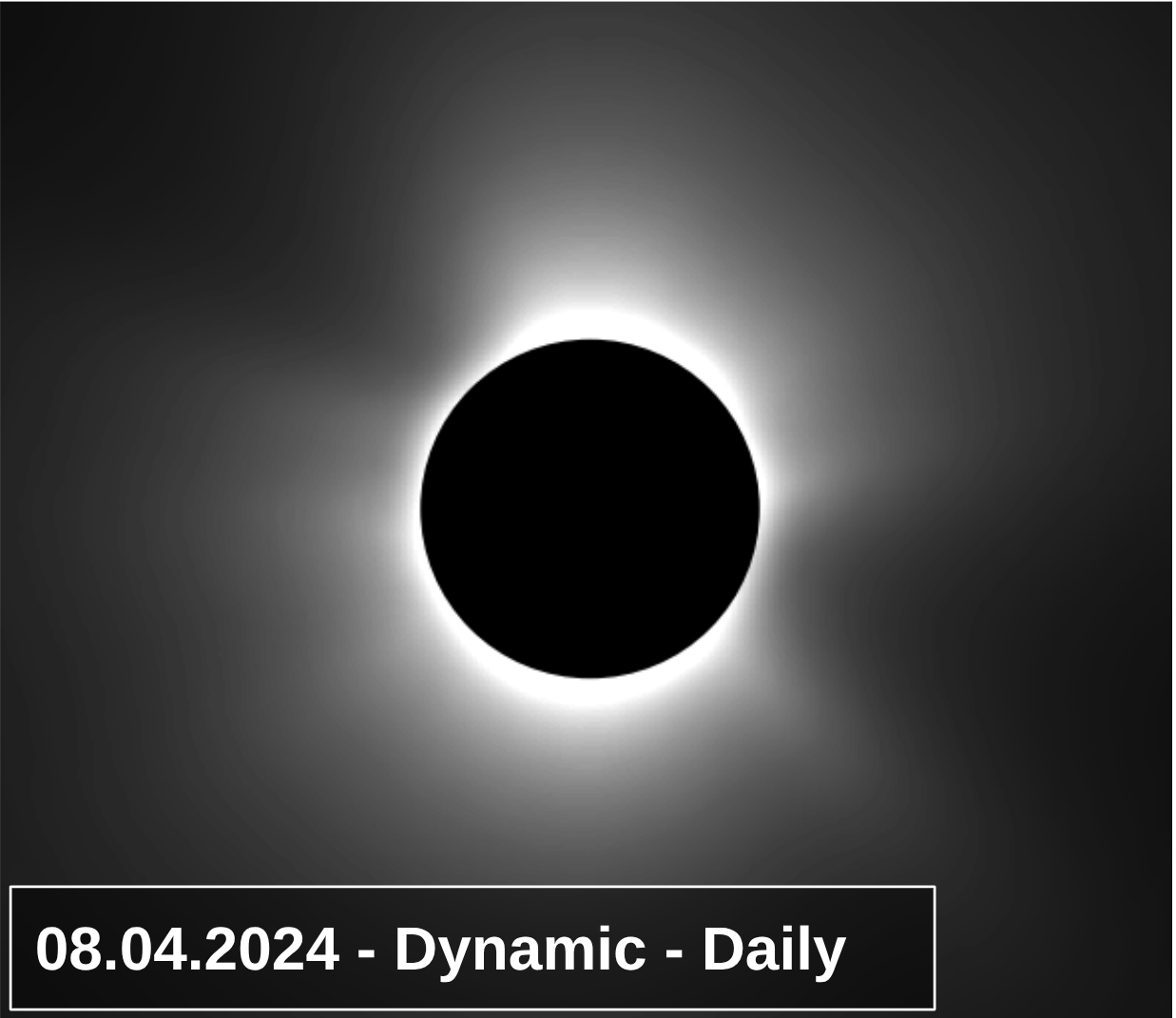}
    \includegraphics[width=0.33\textwidth]{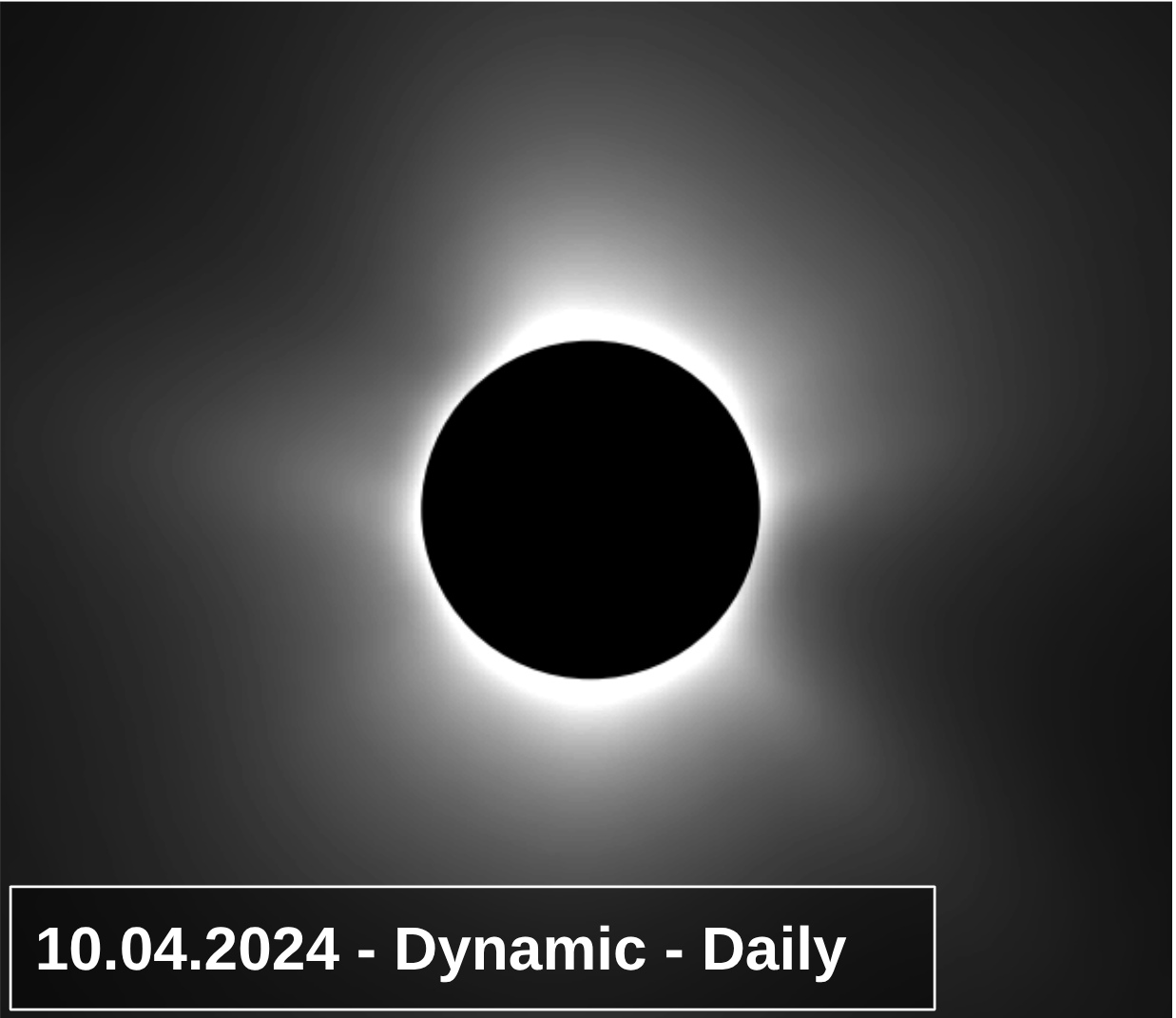}
  \caption{The eclipse and synthetic WL images generated from the COCONUT simulations performed with the magnetogram observed on 08.04.2024 and 10.04.2025. Both COCONUT simulations are performed in the dynamic regime with daily updated magnetograms. {Eclipse image credits}: Eclipse team of Nanjing University - Wu, Sizhe (photographer); Li, Yihua; Huang, Yuhao; Lao, Qinghui; Cheng, Xin; Qu, Zhongquan.}\label{fig:eclipse_image_wli}
\end{figure*}

We can see from the magnetograms that the photospheric magnetic field configuration changes significantly on April 8 compared to April 3. Moreover, in the synoptic magnetograms, the region behind the east limb has not been updated for 14 days. Therefore, after the total solar eclipse, we used a magnetic field configuration on April 10 to model the solar corona and show Earth's field of view as it would appear on April 8. This way, we could see if the additional observations on the east limb would influence the modelled profile. The results are given in Figure~\ref{fig:10Apr_steady_dynamic}. The figures are arranged similarly to the previous figures that illustrate the simulation results. The left figure shows the results for the quasi-steady simulation. The middle and right figures show the results for the dynamic simulations, with the boundaries driven by a daily or hourly cadence for updating magnetic field information. Notice that the streamer directions in all three types of simulations are similar. The west limb is modelled well, similar to previous cases. However, significant differences occur between the modelled magnetic field configuration on the east limb using magnetograms observed before April 8 and those observed on April 10. Notice the tendency of the two clear streamers to converge towards each other in the simulations corresponding to April 8. On April 10, we do not distinguish between the two separate streamers. There is one clear streamer between the previously modelled two streamers, close to region four in our original notation.

\subsection{White-light images and polarised brightness}
\begin{figure*}[hpt!]
   \includegraphics[width=\textwidth]{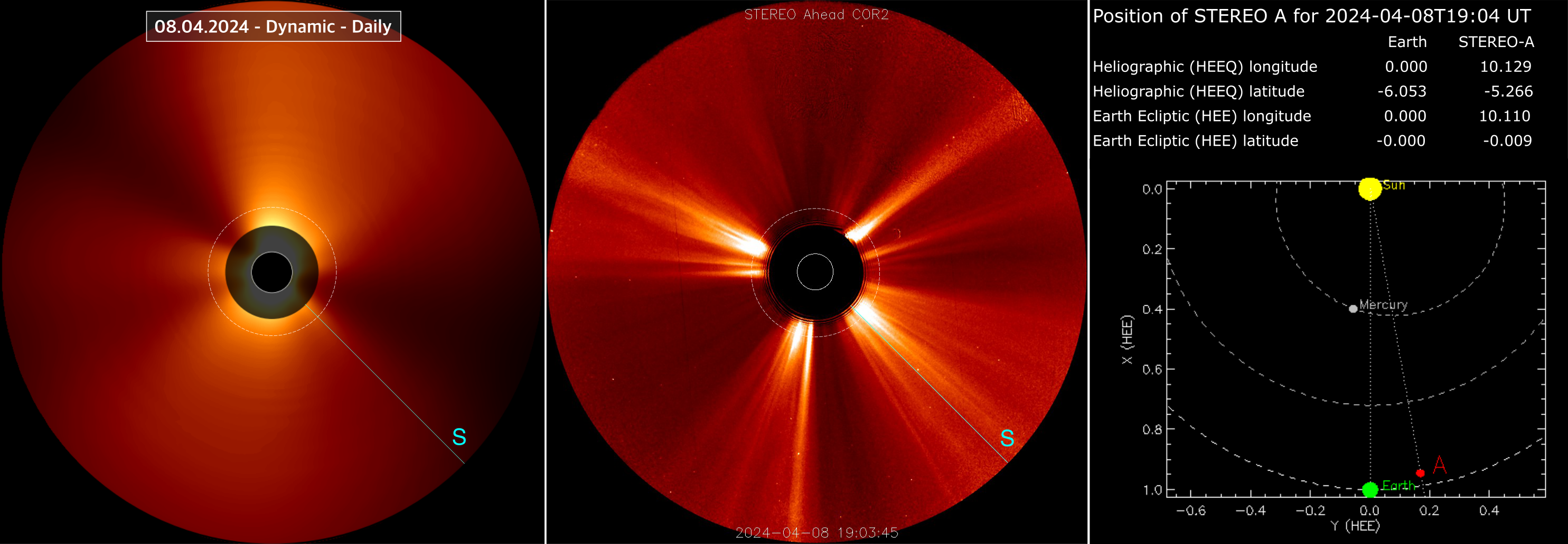}
 \caption{The synthetic {WL }image generated from COCONUT with daily updated magnetograms up to the total solar eclipse {for the STEREO A FOV} and a real STEREO A/COR2 image are given in the first and second panels, respectively. The location of STEREO A with respect to Earth is plotted in the third panel. }\label{fig:cor2_WL_image}
\end{figure*}

In addition to the corona validation technique suggested by \cite{Wagner2022}, which was introduced previously in this section, we generate synthetic WL images to qualitatively compare the modelled data with the observed image. Next, we also compare the modelled polarised brightness to the observed one quantitatively to introduce the most comprehensive assessment of the predictive capabilities of the COCONUT coronal model. 
\subsubsection{Synthetic White-Light images}

The synthetic WL images were generated from COCONUT simulations with the software FoMo. The FoMo code \citep{TVD2016} WL extension {\citep{Sorokina25}} allows for computing WL synthetic data from arbitrary numerical models. The FoMo WL tool operates with data cubes or simulation snapshots in popular data formats. The forward modelling procedure involves interpolating plasma density from the original simulation grid to a Cartesian grid in the observation reference frame. Two rotation angles set the observer's field of view (FOV). The resolution is user-defined. Thus, the synthetic image resolution and the orientation of its FOV can be tuned to the real instrument specifications or set for user analysis. Using the Thomson scattering theory, the interpolated data are converted to radiant intensity, which is integrated along the line of sight (LOS) to produce synthetic observations. FoMo uses open-source software (written in C++ and Python) to process data and can run in parallel on multiple processors on high-performance computing systems. Details on installing and using the FoMo WL tool are available at {GitHub: FoMo}\footnote{\url{https://github.com/TomVeeDee/FoMo}}.

Figure~\ref{fig:eclipse_image_wli} shows the observed image on the left and the synthetic WL images created from COCONUT simulations. The middle panel corresponds to the simulation snapshot driven by the magnetograms up to the total solar eclipse time, and the right panel represents the simulation snapshot corresponding to the eclipse view from Earth, driven by the magnetogram observed up to 10.04.2024. Both simulations are driven in the dynamic regime, with daily updated magnetograms. The synthetic WL images were also generated for the other simulations presented in the previous section. However, here we present only the daily updated regime results, as they yielded slightly better results than steady driving and similar results to the hourly updated simulations. 

We can see two streamers in the west region and a prominent streamer in the south-west region, which are replicated in the synthetic images. Therefore, the west limb is modelled well compared to the real image. We can also identify an enhanced brightness region closer to the South Pole, where significant features were observed in the eclipse image. The streamer is less pronounced in the synthetic images than in the observed eclipse image. We can also see a significant difference between the middle and right panels on the east limb. In the synthetic WL image in the middle panel, we can see two narrow streamers near the equator. In the right panel, we observe these streamers merging and blending into a larger streamer. This is due to new information about the emerging magnetic activity observed in the magnetograms after the eclipse date. The east limb is better modelled by the configuration in the right panel; thus, including the magnetic field information from the back of the Sun improved the modelled image. 

{The total solar eclipse was also modelled by the MAS model \citep{Downs2025}. They began making predictions on March 16, 2025, and also performed post-eclipse simulations on April 15, 2025. The methods and techniques used by the MAS model were significantly different from those used by COCONUT. Namely, they used the data assimilation method for integrating the Polarimetric and Helioseismic Imager (PHI) \citep{Solanki2020} data on the Solar
Orbiter (SolO) into HMI magnetograms, and performed the Surface Flux Transport (SFT) \citep{Caplan2025} (model for creating the inner boundary magnetic field, while COCONUT used GONG input magnetograms filtered with spherical harmonics. Since the total solar eclipse occurred during the earlier development stage of COCONUT, we investigated its modelling capabilities with a simplified approach. We used a low-resolution domain with roughly 380,000 elements, as mentioned before, while the MAS model used $\sim 52$ million elements. Additionally, our input magnetic field was heavily smoothed out, with a strength in the range $\sim \pm20$G, compared to $\sim \pm100$G used in MAS input magnetic field maps. The synthetic WL images generated from MAS simulation results resolve structures in more detail, compared to the COCONUT synthetic WL images, given in Fig.~\ref{fig:eclipse_image_wli}. This is largely due to the resolution used in the simulations and WL generation tools. However, we could see that the strong streamers were modelled reasonably well and fast, each simulation taking $\sim 20$ minutes on 144 CPUs. }

\subsubsection{Polarized brightness}
\begin{figure}[hpt!]
   \includegraphics[width=0.45\textwidth]{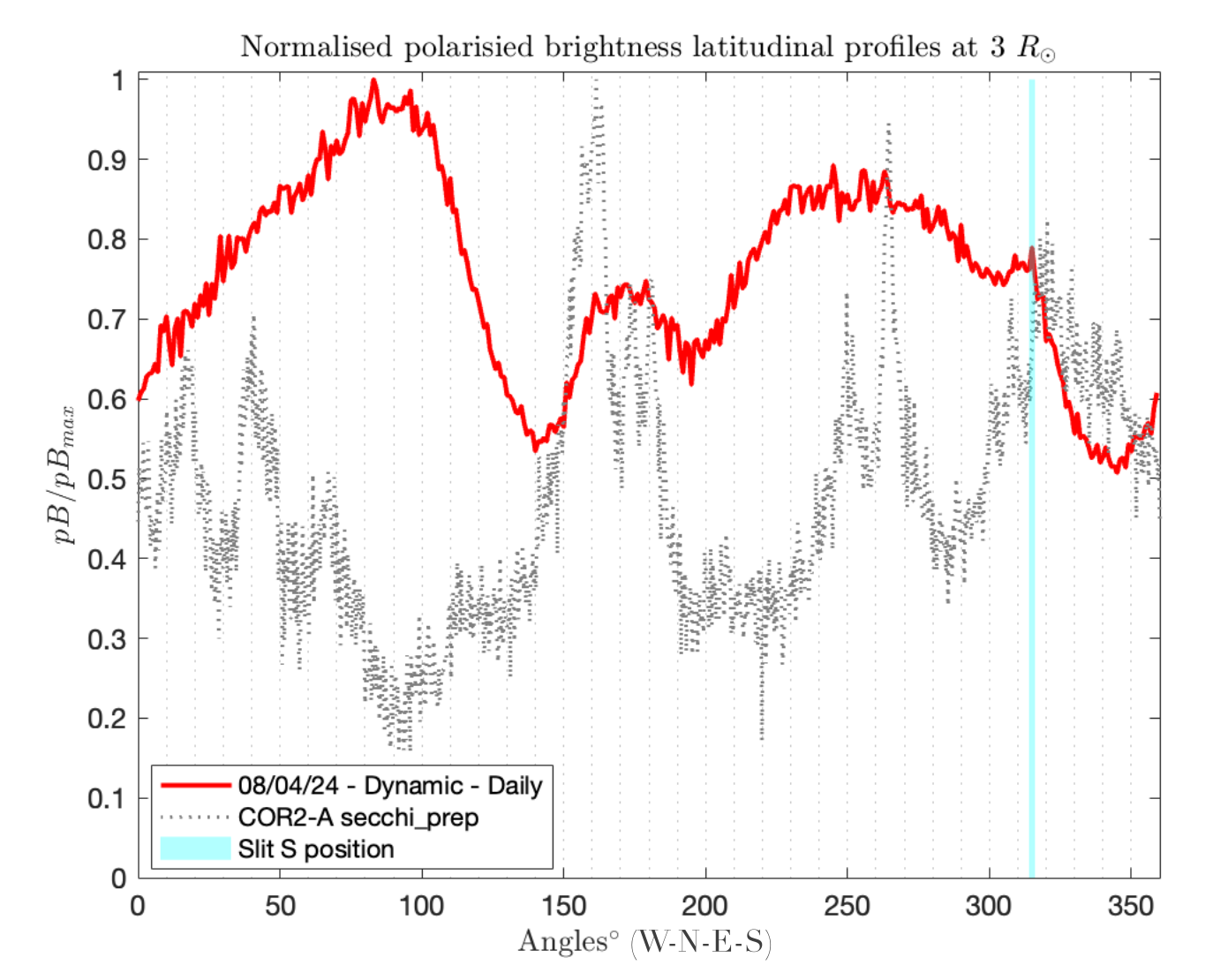}
 \caption{Normalised polarised brightness from COR2-A (data extracted with standard secchi$\_$prep routine) and COCONUT dynamic simulation with daily updated magnetograms. The polarised brightness is given for the entire latitudinal range at a distance of $3\;R_\odot$. The slit position denoted by S in Figure~\ref{fig:cor2_WL_image} is indicated with a cyan vertical line. }\label{fig:normalized_pB_at_3R}
\end{figure}

The modelled corona result is quantitatively compared to the observations by examining the normalised polarised brightness along the radial slit.
Figure~\ref{fig:cor2_WL_image} compares the syntetic WL image produced from COCONUT simulation, driven by daily-updated magnetograms, and the observation by the STEREO A/COR2 \citep{Kaiser2008} around the time of the eclipse on 2024 Apr 8 at 19:03:45. The FOV of the synthetic image matches the one of COR2-A, when the separation angle between the Earth and the coronagraph is 10 degrees (see the right panel in Figure~\ref{fig:cor2_WL_image}). The dashed white circle is drawn at 3~R$_\odot$ and the radial slit is plotted on the figures in cyan along 315$^\circ$. Here and below, the angle of zero degrees corresponds to the westernmost point on the circle and increases counterclockwise. The vertical axis in both images is aligned to the solar rotation axis, and the top of the image corresponds to the Sun's North Pole. In the synthetic image, we hide the area below $2.3\;\text{R}_\odot$ with a semi-transparent mask. The coronal configuration differs slightly from the observed eclipse image from Earth due to the different viewing locations (see the right panel in Figure~\ref{fig:cor2_WL_image}). The North Pole is oversaturated in the COCONUT simulation result, while it appears somewhat dimmed in the observation. The prominent streamer is present on the east limb in the observation and the COCONUT synthetic image. The enhanced region at the south pole is wider in the COCONUT simulation compared to the observation. We observe that the western limb is also well reproduced.

For a more detailed comparison, we consider the shape of the normalised polarised brightness profiles obtained during the observations and after forward modelling. The calibrated observational data for comparison were extracted using the standard IDL/secchi\_prep routine. Figure~\ref{fig:normalized_pB_at_3R} compares the latitudinal profiles at 3~R$_\odot$, shown as a white dashed circle in Figure~\ref{fig:cor2_WL_image}. The normalised polarised brightness profiles also suggest that the North Pole is completely oversaturated compared to the COR2 observations. The streamer between 150$^\circ$ and 200$^\circ$ is present both in the COCONUT simulation and the COR2 image; however, the COCONUT image is undersaturated. The streamer near the South Pole is spread out in the COCONUT simulation, and the profile observed is well recovered after the 315$^\circ$ region. The enhanced profiles near the poles in the simulations could be due to a heavily processed input magnetic map affecting the heating profile. The comparison in Figure~\ref{fig:normalized_pB_at_3R} confirms the conclusion of the qualitative comparison from Figure~\ref{fig:cor2_WL_image}. 
\begin{figure}[hpt!]
   \includegraphics[width=0.5\textwidth]{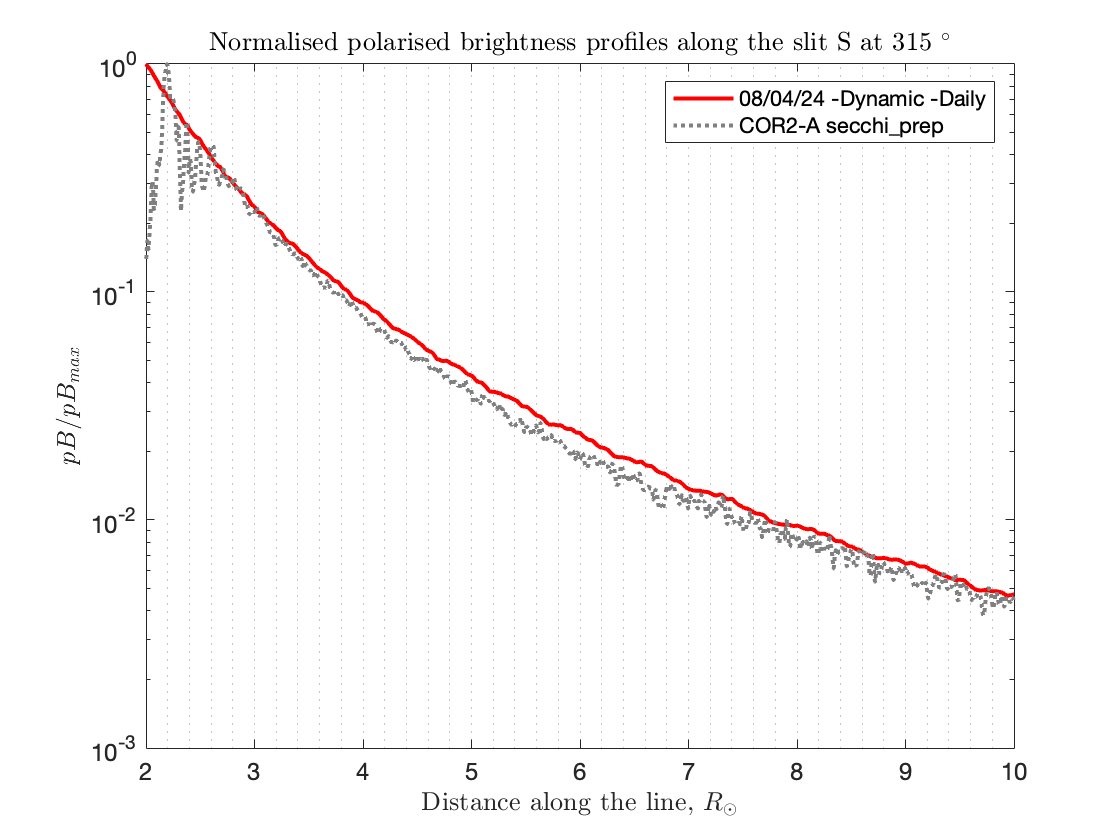}
 \caption{Normalized polarized brightness along the slit denoted by S in Figure~\ref{fig:cor2_WL_image}, corresponding to 315$^\circ$. The from COR2-A (data extracted with standard secchi$\_$prep routine) and COCONUT synthetic polarised brightness is plotted up to 10\;R$_\odot$.} \label{fig:normalized_pB_at_315_degrees}
\end{figure}

Figure~\ref{fig:normalized_pB_at_315_degrees} compares the radial profiles along the line S at 315$^\circ$, shown as a cyan line in Figure~\ref{fig:cor2_WL_image}. Here, we can see that the radial profiles of the polarised brightness in the observed image and the COCONUT simulation agree well. 

\section{Conclusions} \label{sec:conclusions}
The total solar eclipse on April 8, 2024, is used to validate the COCONUT coronal model. Previously, only the simple, polytropic version of the COCONUT model was used to assess the low corona configuration of the Sun. This study utilises the full MHD version of COCONUT, which includes the source and sink terms, providing a more detailed description of the heating, thermal conduction, and radiative loss processes near the Sun. Due to the high solar activity during the eclipse, highly processed maps were utilised with a low-resolution grid for the computations. The total solar corona predictions started on March 21, 2024, and were run in different regimes. First, the predictions were performed in the quasi-steady mode. Steady simulations were run daily before the total solar eclipse, and their evolution was monitored over 19 days.

Each of these simulations took $\sim20$ minutes on 4 nodes with 36 cores each on the Genius cluster of the Flemish supercomputer centre. Then, after the eclipse, time-dependent simulations were performed using the same daily updated magnetograms as those used for the quasi-steady predictions. The magnetograms were interpolated and fed into the COCONUT simulation as time-evolving inner boundary conditions. The last regime we investigated considered the time-dependent, dynamic simulations with hourly updated magnetograms. This way, we investigated the effect of dynamic driving of the boundary and the cadence of the inner boundary. 

We utilised various techniques available in the literature to validate the obtained results. Since the input magnetic maps were highly processed and the domain resolution was low, using multiple validation techniques would be beneficial. First, we examined the configuration of the magnetic streamers in the modelled solar corona and overlaid the magnetic field lines onto the total solar eclipse image. This was done for both quasi-steady and dynamic simulations with low and high cadence. We observed that the most similar configuration was achieved in the dynamic simulations. However, increasing the cadence of the inner boundary conditions did not notably improve the results, likely due to the strong preprocessing of the maps. 

We also performed simulations that included the magnetograms observed after the total solar eclipse. We could see that the east limb was modelled differently due to the updated magnetic field information from the back of the Sun. Based on the comparison of the magnetic field configuration of the solar corona, we used only dynamic simulations driven by a daily cadence for the following validation techniques. This implied the synthetic WL images and normalised polarised brightness profiles. From this analysis, we could see that the prominent streamers were reconstructed well on the west limb, and the streamer's location on the east limb was affected by including the information from the back of the Sun. The normalised polarised brightness image at 3\;R$_\odot$ confirmed the comparison of the latitudinal distribution of the enhanced regions in the observed and modelled images. The normalised polarised brightness along 315$^\circ$ radial slit showed that the profiles of the observed and modelled corona are in good agreement. 

Predicting the total solar eclipse of 2024 was challenging due to high solar activity and the presence of strong magnetic field regions on the solar surface. Using multiple validating techniques was essential to assess the model's performance and determine the role of processing the input data in the obtained results. Even though the pronounced streamers were reconstructed in the model, the main limitation of the model remains the lack of a detailed magnetic field configuration at the inner boundary. Improving the robustness of the COCONUT model is the next step in its development. For future work, we intend to investigate the relationship between the processed magnetic field and the generated synthetic images while light in the simulations. Additionally, we intend to perform coronal modelling with a less processed magnetogram on a higher resolution computational domain to reconstruct the observed features in more detail and avoid smoothing them over vast regions.

\begin{acknowledgements}
This research has received funding from the European Union’s Horizon 2020 research and innovation programme under grant agreement No 870405 (EUHFORIA 2.0) and the ESA project "Heliospheric Modelling Techniques“ (Contract No. 4000133080/20/NL/CRS).
These results were also obtained in the framework of the projects C16/24/010  (C1 project Internal Funds KU Leuven), AFOSR FA9550-18-1-0093, G0B5823N and G002523N  (FWO-Vlaanderen), and SIDC Data Exploitation (ESA Prodex-12).
The computational resources and services used in this work were provided by the VSC-Flemish Supercomputer Center, funded by the Research Foundation Flanders (FWO) and the Flemish Government-Department EWI.

SP also acknowledges funding by the European Union. Views and opinions expressed are, however, those of the author(s) only and do not necessarily reflect those of the European Union or ERCEA. Neither the European Union nor the granting authority can be held responsible for them. This project (Open SESAME) has received funding under the Horizon Europe programme (ERC-AdG agreement No 101141362).

The STEREO/SECCHI data used here are produced by an international consortium of the Naval Research Laboratory (USA), Lockheed Martin Solar and Astrophysics Laboratory (USA), NASA Goddard Space Flight Center (USA), Rutherford Appleton Laboratory (UK), University of Birmingham (UK), Max-Planck-Institut für Sonnensystemforschung (Germany), Centre Spatial de Liège (Belgium), Institut d’Optique Théorique et Appliqué (France), and Institut d’Astrophysique Spatiale (France).
\end{acknowledgements}
\bibliographystyle{aa}
\bibliography{bibliography}

\begin{thebibliography}{35}
\expandafter\ifx\csname natexlab\endcsname\relax\def\natexlab#1{#1}\fi

\bibitem[{{Baratashvili} {et~al.}(2024){Baratashvili}, {Brchnelova}, {Linan}, {Lani}, \& {Poedts}}]{Baratashvili2024C}
{Baratashvili}, T., {Brchnelova}, M., {Linan}, L., {Lani}, A., \& {Poedts}, S. 2024, \aap, 690, A184

\bibitem[{{Baratashvili} {et~al.}(2025){Baratashvili}, {Popescu Braileanu}, {Bacchini}, {Keppens}, \& {Poedts}}]{Baratashvili2025B}
{Baratashvili}, T., {Popescu Braileanu}, B., {Bacchini}, F., {Keppens}, R., \& {Poedts}, S. 2025, \aap, 694, A306

\bibitem[{{Brchnelova} {et~al.}(2022{\natexlab{a}}){Brchnelova}, {Ku\'zma}, {Perri}, {Baratishvili}, {Zhang}, {Lani}, \& {Poedts}}]{Brchnelova2022}
{Brchnelova}, M., {Ku\'zma}, B., {Perri}, B., {et~al.} 2022{\natexlab{a}}, \apjs, accepted

\bibitem[{{Brchnelova} {et~al.}(2022{\natexlab{b}}){Brchnelova}, {Zhang}, {Leitner}, {Perri}, {Lani}, \& {Poedts}}]{Brchnelova2022b}
{Brchnelova}, M., {Zhang}, F., {Leitner}, P., {et~al.} 2022{\natexlab{b}}, Journal of Plasma Physics, 88, 905880205

\bibitem[{{Caplan} {et~al.}(2025){Caplan}, {Stulajter}, {Linker}, {Downs}, {Upton}, {Jha}, {Attie}, {Arge}, \& {Henney}}]{Caplan2025}
{Caplan}, R.~M., {Stulajter}, M.~M., {Linker}, J.~A., {et~al.} 2025, \apjs, 278, 24

\bibitem[{{Chorin}(1997)}]{chorin1997}
{Chorin}, A.~J. 1997, Journal of Computational Physics, 135, 118

\bibitem[{{Dedner} {et~al.}(2002){Dedner}, {Kemm}, {Kr{\"o}ner}, {Munz}, {Schnitzer}, \& {Wesenberg}}]{Dedner2002}
{Dedner}, A., {Kemm}, F., {Kr{\"o}ner}, D., {et~al.} 2002, Journal of Computational Physics, 175, 645

\bibitem[{{Downs} {et~al.}(2025){Downs}, {Linker}, {Caplan}, {Mason}, {Riley}, {Davidson}, {Reyes}, {Palmerio}, {Lionello}, {Turtle}, {Ben-Nun}, {Stulajter}, {Titov}, {T{\"o}r{\"o}k}, {Upton}, {Attie}, {Jha}, {Arge}, {Henney}, {Valori}, {Strecker}, {Calchetti}, {Germerott}, {Hirzberger}, {Su{\'a}rez}, {Rodr{\'\i}guez}, {Solanki}, {Cheng}, \& {Wu}}]{Downs2025}
{Downs}, C., {Linker}, J.~A., {Caplan}, R.~M., {et~al.} 2025, Science, 388, 1306

\bibitem[{{Kaiser} {et~al.}(2008){Kaiser}, {Kucera}, {Davila}, {St.~Cyr}, {Guhathakurta}, \& {Christian}}]{Kaiser2008}
{Kaiser}, M.~L., {Kucera}, T.~A., {Davila}, J.~M., {et~al.} 2008, \ssr, 136, 5

\bibitem[{Kimpe {et~al.}(2005)Kimpe, Lani, Quintino, Poedts, \& Vandewalle}]{Kimpe2005}
Kimpe, D., Lani, A., Quintino, T., Poedts, S., \& Vandewalle, S. 2005, in Recent Advances in Parallel Virtual Machine and Message Passing Interface, ed. B.~Di~Martino, D.~Kranzlm{\"u}ller, \& J.~Dongarra (Berlin, Heidelberg: Springer Berlin Heidelberg), 520--527

\bibitem[{{Ku{\'z}ma} {et~al.}(2023){Ku{\'z}ma}, {Brchnelova}, {Perri}, {Baratashvili}, {Zhang}, {Lani}, \& {Poedts}}]{Kuzma2023}
{Ku{\'z}ma}, B., {Brchnelova}, M., {Perri}, B., {et~al.} 2023, \apj, 942, 31

\bibitem[{Lani {et~al.}(2005)Lani, Quintino, Kimpe, Deconinck, Vandewalle, \& Poedts}]{Lani2005}
Lani, A., Quintino, T., Kimpe, D., {et~al.} 2005, in Computational Science -- ICCS 2005, ed. V.~S. Sunderam, G.~D. van Albada, P.~M.~A. Sloot, \& J.~J. Dongarra (Berlin, Heidelberg: Springer Berlin Heidelberg), 279--286

\bibitem[{Lani {et~al.}(2006)Lani, Quintino, Kimpe, Deconinck, Vandewalle, \& Poedts}]{Lani2006}
Lani, A., Quintino, T., Kimpe, D., {et~al.} 2006, Scientific Programming, 14

\bibitem[{Lani {et~al.}(2013)Lani, Villedieu, Bensassi, Kapa, Vymazal, Yalim, \& Panesi}]{Lani2013}
Lani, A., Villedieu, N., Bensassi, K., {et~al.} 2013, in AIAA 2013-2589, 21th AIAA CFD Conference, San Diego (CA)

\bibitem[{{Lani} {et~al.}(2014){Lani}, {Yalim}, \& {Poedts}}]{Lani2014}
{Lani}, A., {Yalim}, M.~S., \& {Poedts}, S. 2014, Computer Physics Communications, 185, 2538

\bibitem[{{Miki{\'c}} {et~al.}(1999){Miki{\'c}}, {Linker}, {Schnack}, {Lionello}, \& {Tarditi}}]{Mikic1999}
{Miki{\'c}}, Z., {Linker}, J.~A., {Schnack}, D.~D., {Lionello}, R., \& {Tarditi}, A. 1999, Physics of Plasmas, 6, 2217

\bibitem[{Mikić \& Linker(1996)}]{MikicLinker1996}
Mikić, Z. \& Linker, J.~A. 1996, AIP Conference Proceedings, 382, 104

\bibitem[{{Perri} {et~al.}(2022){Perri}, {Leitner}, {Brchnelova}, {Baratishvili}, Ku{\'{z}}ma, {Zhang}, {Lani}, \& {Poedts}}]{Perri2022}
{Perri}, B., {Leitner}, P., {Brchnelova}, M., {et~al.} 2022, \apj, 936

\bibitem[{{Pinto} \& {Rouillard}(2017)}]{Pinto2017}
{Pinto}, R.~F. \& {Rouillard}, A.~P. 2017, \apj, 838, 89

\bibitem[{{Pomoell} \& {Poedts}(2018)}]{Pomoell2018}
{Pomoell}, J. \& {Poedts}, S. 2018, Journal of Space Weather and Space Climate, 8, A35

\bibitem[{{R{\'e}ville} {et~al.}(2016){R{\'e}ville}, {Folsom}, {Strugarek}, \& {Brun}}]{reville2016}
{R{\'e}ville}, V., {Folsom}, C.~P., {Strugarek}, A., \& {Brun}, A.~S. 2016, \apj, 832, 145

\bibitem[{{R{\'e}ville} {et~al.}(2020){R{\'e}ville}, {Velli}, {Panasenco}, {Tenerani}, {Shi}, {Badman}, {Bale}, {Kasper}, {Stevens}, {Korreck}, {Bonnell}, {Case}, {de Wit}, {Goetz}, {Harvey}, {Larson}, {Livi}, {Malaspina}, {MacDowall}, {Pulupa}, \& {Whittlesey}}]{reville2020}
{R{\'e}ville}, V., {Velli}, M., {Panasenco}, O., {et~al.} 2020, \apjs, 246, 24

\bibitem[{{Schatten} {et~al.}(1969){Schatten}, {Wilcox}, \& {Ness}}]{Schatten1969}
{Schatten}, K.~H., {Wilcox}, J.~M., \& {Ness}, N.~F. 1969, \solphys, 6, 442

\bibitem[{{Solanki} {et~al.}(2020){Solanki}, {del Toro Iniesta}, {Woch}, {Gandorfer}, {Hirzberger}, {Alvarez-Herrero}, {Appourchaux}, {Mart{\'\i}nez Pillet}, {P{\'e}rez-Grande}, {Sanchis Kilders}, {Schmidt}, {G{\'o}mez Cama}, {Michalik}, {Deutsch}, {Fernandez-Rico}, {Grauf}, {Gizon}, {Heerlein}, {Kolleck}, {Lagg}, {Meller}, {M{\"u}ller}, {Sch{\"u}hle}, {Staub}, {Albert}, {Alvarez Copano}, {Beckmann}, {Bischoff}, {Busse}, {Enge}, {Frahm}, {Germerott}, {Guerrero}, {L{\"o}ptien}, {Meierdierks}, {Oberdorfer}, {Papagiannaki}, {Ramanath}, {Schou}, {Werner}, {Yang}, {Zerr}, {Bergmann}, {Bochmann}, {Heinrichs}, {Meyer}, {Monecke}, {M{\"u}ller}, {Sperling}, {{\'A}lvarez Garc{\'\i}a}, {Aparicio}, {Balaguer Jim{\'e}nez}, {Bellot Rubio}, {Cobos Carracosa}, {Girela}, {Hern{\'a}ndez Exp{\'o}sito}, {Herranz}, {Labrousse}, {L{\'o}pez Jim{\'e}nez}, {Orozco Su{\'a}rez}, {Ramos}, {Barandiar{\'a}n}, {Bastide}, {Campuzano}, {Cebollero}, {D{\'a}vila}, {Fern{\'a}ndez-Medina}, {Garc{\'\i}a Parejo}, {Garranzo-Garc{\'\i}a}, {Laguna},
  {Mart{\'\i}n}, {Navarro}, {N{\'u}{\~n}ez Peral}, {Royo}, {S{\'a}nchez}, {Silva-L{\'o}pez}, {Vera}, {Villanueva}, {Fourmond}, {de Galarreta}, {Bouzit}, {Hervier}, {Le Clec'h}, {Szwec}, {Chaigneau}, {Buttice}, {Dominguez-Tagle}, {Philippon}, {Boumier}, {Le Cocguen}, {Baranjuk}, {Bell}, {Berkefeld}, {Baumgartner}, {Heidecke}, {Maue}, {Nakai}, {Scheiffelen}, {Sigwarth}, {Soltau}, {Volkmer}, {Blanco Rodr{\'\i}guez}, {Domingo}, {Ferreres Sabater}, {Gasent Blesa}, {Rodr{\'\i}guez Mart{\'\i}nez}, {Osorno Caudel}, {Bosch}, {Casas}, {Carmona}, {Herms}, {Roma}, {Alonso}, {G{\'o}mez-Sanjuan}, {Piqueras}, {Torralbo}, {Fiethe}, {Guan}, {Lange}, {Michel}, {Bonet}, {Fahmy}, {M{\"u}ller}, \& {Zouganelis}}]{Solanki2020}
{Solanki}, S.~K., {del Toro Iniesta}, J.~C., {Woch}, J., {et~al.} 2020, \aap, 642, A11

\bibitem[{{Sorokina} {et~al.}(2025){Sorokina}, {Van Doorsselaere}, {Lloveras}, {Linan}, \& {Poedts}}]{Sorokina25}
{Sorokina}, D., {Van Doorsselaere}, T., {Lloveras}, D.~G., {Linan}, L., \& {Poedts}, S. 2025, A\&A, 701, A166

\bibitem[{{Usmanov} {et~al.}(2014){Usmanov}, {Goldstein}, \& {Matthaeus}}]{Usmanov2014}
{Usmanov}, A.~V., {Goldstein}, M.~L., \& {Matthaeus}, W.~H. 2014, \apj, 788, 43

\bibitem[{{Usmanov} {et~al.}(2018){Usmanov}, {Matthaeus}, {Goldstein}, \& {Chhiber}}]{Usmanov20189}
{Usmanov}, A.~V., {Matthaeus}, W.~H., {Goldstein}, M.~L., \& {Chhiber}, R. 2018, \apj, 865, 25

\bibitem[{{van der Holst} {et~al.}(2014){van der Holst}, {Sokolov}, {Meng}, {Jin}, {Manchester}, {T{\'o}th}, \& {Gombosi}}]{Vanderholst2014}
{van der Holst}, B., {Sokolov}, I.~V., {Meng}, X., {et~al.} 2014, \apj, 782, 81

\bibitem[{Van~Doorsselaere {et~al.}(2016)Van~Doorsselaere, Antolin, Yuan, Reznikova, \& Magyar}]{TVD2016}
Van~Doorsselaere, T., Antolin, P., Yuan, D., Reznikova, V., \& Magyar, N. 2016, Frontiers in Astronomy and Space Sciences, 3:4

\bibitem[{{Wagner} {et~al.}(2022){Wagner}, {Asvestari}, {Temmer}, {Heinemann}, \& {Pomoell}}]{Wagner2022}
{Wagner}, A., {Asvestari}, E., {Temmer}, M., {Heinemann}, S.~G., \& {Pomoell}, J. 2022, \aap, 657, A117

\bibitem[{{Wang} {et~al.}(2025{\natexlab{a}}){Wang}, {Poedts}, {Lani}, {Brchnelova}, {Baratashvili}, {Linan}, {Zhang}, {Hou}, \& {Zhou}}]{Wang2025}
{Wang}, H.~P., {Poedts}, S., {Lani}, A., {et~al.} 2025{\natexlab{a}}, \aap, 694, A234

\bibitem[{{Wang} {et~al.}(2025{\natexlab{b}}){Wang}, {Poedts}, {Lani}, {Linan}, {Baratashvili}, {Zhang}, {Sorokina}, {Jeong}, {Li}, {Najafi-Ziyazi}, \& {Schmieder}}]{wang2025COCONUTMayEvent}
{Wang}, H.~P., {Poedts}, S., {Lani}, A., {et~al.} 2025{\natexlab{b}}, \aap, 702, A37

\bibitem[{{Wang} {et~al.}(2025{\natexlab{c}}){Wang}, {Yang}, {Poedts}, {Lani}, {Zhou}, {Gao}, {Linan}, {Lv}, {Baratashvili}, {Guo}, {Lin}, {Su}, {Li}, {Zhang}, {Wei}, {Yang}, {Li}, {Ma}, {Husidic}, H.-J., {Mahdi}, {Wang}, \& {Schmieder}}]{wang2025sipifvmtimeevolvingcoronalmodel}
{Wang}, H.~P., {Yang}, L.~P., {Poedts}, S., {et~al.} 2025{\natexlab{c}}, \apjs, 278, 59

\bibitem[{{Yalim} {et~al.}(2011){Yalim}, {Vanden Abeele}, {Lani}, {Quintino}, \& {Deconinck}}]{Yalim20211}
{Yalim}, M.~S., {Vanden Abeele}, D., {Lani}, A., {Quintino}, T., \& {Deconinck}, H. 2011, Journal of Computational Physics, 230, 6136

\bibitem[{{Zhukov} {et~al.}(2025){Zhukov}, {Thizy}, {Galano}, {Bourgoignie}, {Dolla}, {Jean}, {Nicula}, {Shestov}, {Galy}, {Rougeot}, {Versluys}, {Zender}, {Lamy}, {Fineschi}, {Gunar}, {Inhester}, {Mierla}, {Rudawy}, {Tsinganos}, {Koutchmy}, {Howard}, {Peter}, {Vives}, {Abbo}, {Aime}, {Aleksiejuk}, {Baran}, {Bak-Steslicka}, {Bemporad}, {Berghmans}, {Besliu-Ionescu}, {Buckley}, {Buiu}, {Capobianco}, {Cimoch}, {DHuys}, {Dziezyc}, {Fleury-Frenette}, {Gibson}, {Giordano}, {Golub}, {Grochowski}, {Heinzel}, {Hermans}, {Jacobs}, {Jejcic}, {Kranitis}, {Landini}, {Loreggia}, {Magdalenic}, {Maia}, {Marque}, {Melich}, {Morawski}, {Mosdorf}, {Noce}, {Orleanski}, {Paschalis}, {Peresty}, {Rodriguez}, {Seaton}, {Short}, {Simar}, {Steslicki}, {Sorensen}, {Terrasa}, {Van Vooren}, {Verstringe}, \& {Zangrilli}}]{Zhukov2025}
{Zhukov}, A.~N., {Thizy}, C., {Galano}, D., {et~al.} 2025, arXiv e-prints, arXiv:2509.00253

\end{thebibliography}
\begin{appendix}
\section{Areas of interest in the input maps}\label{maps_appendix}
\begin{figure}[hpt!]
\centering
    \includegraphics[width=0.5\textwidth]{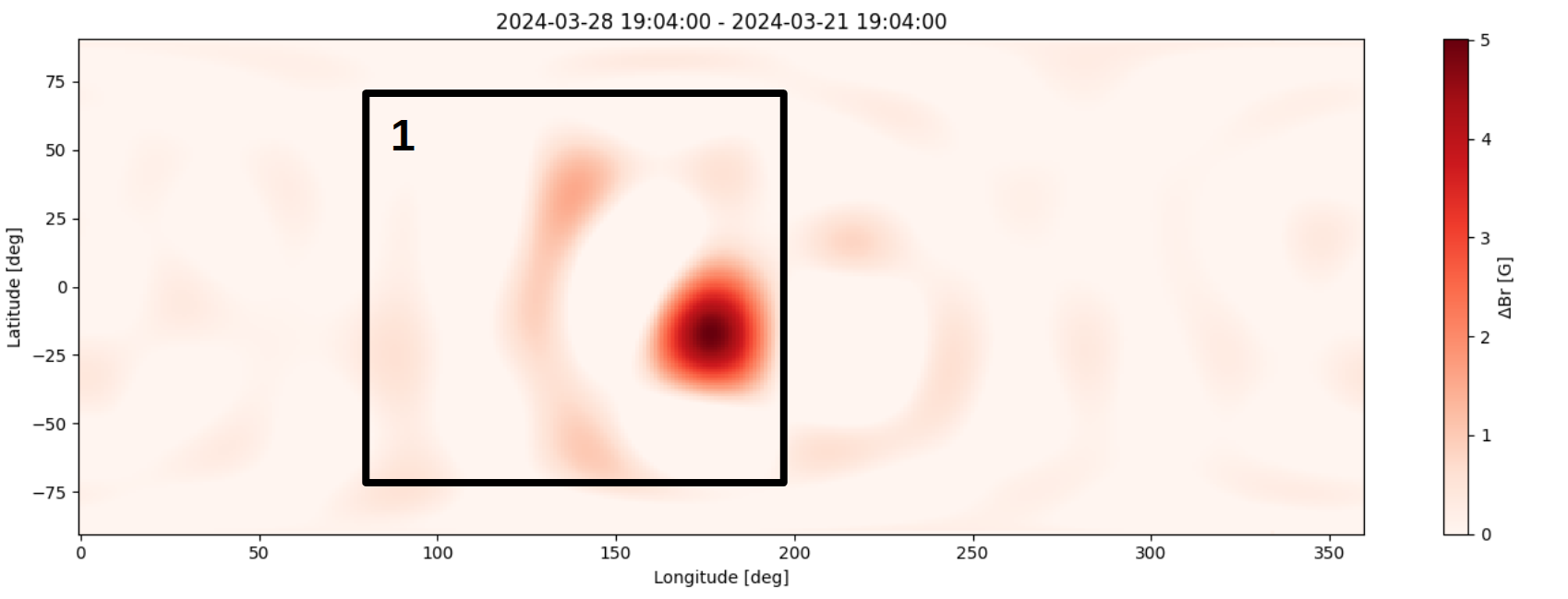}
    \hfill
    \includegraphics[width=0.5\textwidth]{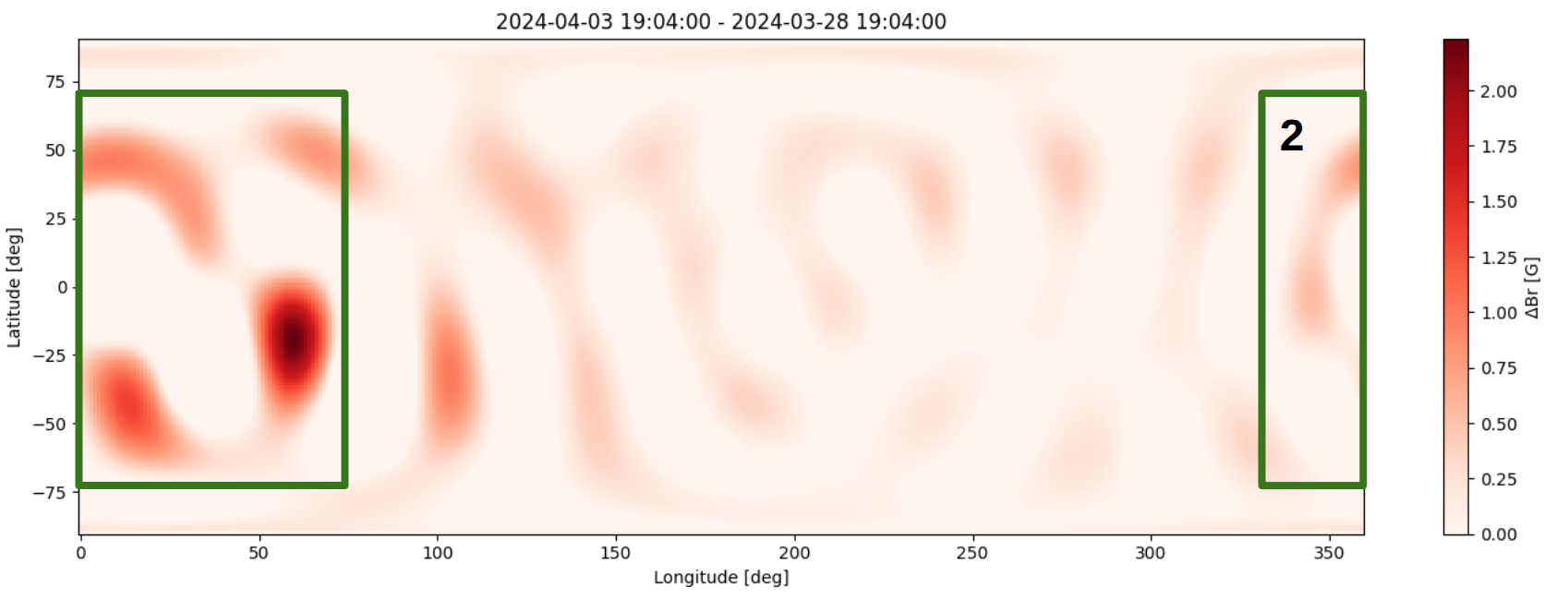}
    \hfill
    \includegraphics[width=0.5\textwidth]{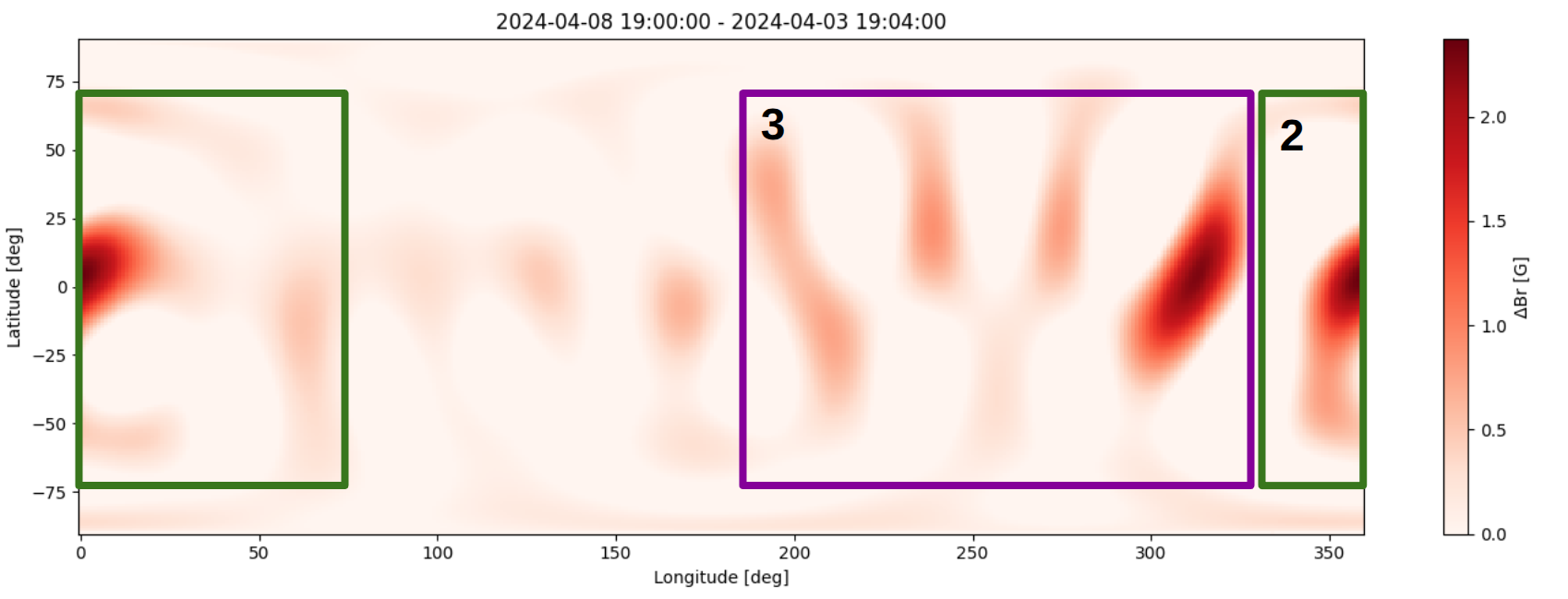}
    \hfill
  \caption{The difference between the smoothed input magnetic field data between the consecutive dates shown in Figure~1 of the paper. The title indicates the dates between which the difference was calculated. }\label{fig:Br_diff_images}
\end{figure}

Figure~\ref{fig:Br_diff_images} shows the calculated difference between the processed radial magnetic field values of the consecutive input magnetic field data given in Figure~\ref{fig:input_magnetograms}. The same boxes as in Figure~\ref{fig:input_magnetograms} are plotted on top of the difference plots. Box 1 shows the difference between the input magnetogram observed on March 28 and March 21. Figure~\ref{fig:Br_diff_images} shows that the strongest differences are present near the right edge of the box, where the magnetic field is also the strongest in the input magnetogram shown in Figure~\ref{fig:input_magnetograms}. However, when considering the structure of the magnetic field configuration, we notice that near the left edge of the box, the slightly enhanced differences indicated lead to a change in the magnetic field structure in the input magnetogram. Thus, we also included this region in this box, rather than limiting it only to the strongest part. As a result, we considered both the magnetic field configuration in the input maps and the calculated differences between consecutive maps to identify the three areas of interest where the most change occurs over time. 
\end{appendix}

\end{document}